\documentclass[graybox]{svmult}

\usepackage[utf8]{inputenc}
\usepackage[T1]{fontenc}
\usepackage{lmodern}
\usepackage{mathptmx}
\usepackage{helvet}
\usepackage{courier}
\usepackage{type1cm}
\usepackage{makeidx}     
\usepackage{graphicx}      
\usepackage{multicol}     
\usepackage[bottom]{footmisc}
\usepackage{enumerate}

\newcommand{\ket}[1]{\ensuremath{\left|{#1}\right\rangle}}
\newcommand{\bra}[1]{\ensuremath{\left\langle{#1}\right|}}

\newcommand{\pare}[1]{\left(#1\right)}
\newcommand{\proj}[2]{\left\vert#1\rangle\langle#2\right\vert}
\makeindex

\begin{document}

\title*{Quantum simulations \\ with circuit quantum electrodynamics}
\author{Guillermo Romero, Enrique Solano, and Lucas Lamata}
\institute{Guillermo Romero \at Departamento de F\'isica, Universidad de Santiago de Chile (USACH), Avenida Ecuador 3493, 9170124 Santiago, Chile, \email{guillermo.romero@usach.cl}
\and Enrique Solano \at Department of Physical Chemistry, University of the Basque Country UPV/EHU, Apartado 644, E-48080 Bilbao, Spain
\at IKERBASQUE, Basque Foundation for Science, Maria Diaz de Haro 3, 48013 Bilbao, Spain, \email{enr.solano@gmail.com}
\and Lucas Lamata \at Department of Physical Chemistry, University of the Basque Country UPV/EHU, Apartado 644, E-48080 Bilbao, Spain, \email{lucas.lamata@gmail.com}}

\maketitle

\abstract*{Each chapter should be preceded by an abstract (10--15 lines long) that summarizes the content. The abstract will appear \textit{online} at \url{www.SpringerLink.com} and be available with unrestricted access. This allows unregistered users to read the abstract as a teaser for the complete chapter. As a general rule the abstracts will not appear in the printed version of your book unless it is the style of your particular book or that of the series to which your book belongs.
Please use the 'starred' version of the new Springer \texttt{abstract} command for typesetting the text of the online abstracts (cf. source file of this chapter template \texttt{abstract}) and include them with the source files of your manuscript. Use the plain \texttt{abstract} command if the abstract is also to appear in the printed version of the book.}

\abstract{Superconducting circuits have become a leading quantum technology for testing fundamentals of quantum mechanics and for the implementation of advanced quantum information protocols. In this chapter, we revise the basic concepts of circuit network theory and circuit quantum electrodynamics for the sake of digital and analog quantum simulations of quantum field theories, relativistic quantum mechanics, and many-body physics, involving fermions and bosons. Based on recent improvements in scalability, controllability, and measurement, superconducting circuits can be considered as a promising quantum platform for building scalable digital and analog quantum simulators, enjoying unique and distinctive properties when compared to other advanced platforms as trapped ions, quantum photonics and optical lattices.}

\section{Introduction}
Nowadays, the field of quantum simulations \cite{Kendon2010,CiracZoller2010,Hauke2012,Monroe2013,Buluta2009,Ashhab2013,Jaksch2013} is one of the most active in quantum information science. Following the original idea by Richard Feynman \cite{Feynman82}, and its subsequent development by Seth Lloyd \cite{Lloyd96}, this field has experienced a significant growth in the last decade. The motivation is the fact that a large quantum system cannot be efficiently simulated with a classical computer due to the exponential growth of the Hilbert space dimension with the number of quantum subsystems. On the other hand, it should be feasible to reproduce the dynamics of quantum systems making use of other, controllable, quantum platforms, which constitute a quantum simulator. Superconducting circuits \cite {DMLesHouches,WCNature,DSScience} and circuit quantum electrodynamics (QED) \cite{Blais2004,Wallraff2004,Chiorescu2004} represent prime candidates to implement a quantum simulator because of their scalability and controllability. There have been already some proposals for quantum simulations in superconducting qubits, as is the case of the quantum simulation of Anderson and Kondo lattices \cite{Ripoll2008}, sudden phase switching in a superconducting qubit array \cite{Tian2010}, molecular collisions \cite{Pritchett}, quantum phases \cite{Zhang2014}, Holstein polarons \cite{Mei2013}, and quantum magnetism \cite{Viehmann2013,Kurcz2014}. Moreover, three pioneering experiments on digital quantum simulators of fermions \cite{Barends2015} and spins \cite{Salathe2015,Barends2015b} have been performed.

In this chapter, we introduce the main concepts of superconducting circuits and circuit QED, and their performance as a quantum simulator. In particular, in Sec.~\ref{sec:circuitnetwork}, we will describe the circuit network theory and the Hamiltonian description of a quantum circuit. In Sec.~\ref{sec:circuitQED}, we provide an introduction to circuit QED and cavity-cavity coupling mechanism, pointing out the coupling regimes of light-matter interaction as building blocks for circuit QED lattices. In Sec.~\ref{sec:AnalogQS}, we discuss the analog quantum simulation of many-body states of light and relativistic phenomena. In addition, in Sec.~\ref{sec:dig1}, we present our recent proposals of digital quantum simulations in circuit QED, such as spin chains and quantum field theories. Finally, in Sec.~\ref{sec:Conc}, we present our concluding remarks.           

\section{Circuit Network Theory} 
\label{sec:circuitnetwork}
Nowadays, integrated quantum circuits \cite{DMLesHouches,WCNature,DSScience} have become a leading technology for quantum information processing and quantum simulations. These devices present noticeable features such as scalability, controllability, and tunable physical parameters which in turn allow us to engineer complex Hamiltonians. In this sense, it is important to understand how an integrated circuit shows its quantum nature, and how to design \textit{two-level systems} or \textit{qubits}. To achieve it, there are three features that one should point out: (i) ultra-low dissipation provided by \textit{superconductivity}, (ii) ultra-low noise reached by \textit{low temperatures}, and (iii) nonlinear, non dissipative elements implemented by Josephson junctions. 

\begin{figure}[t]
\sidecaption[t]
\includegraphics[scale=.55]{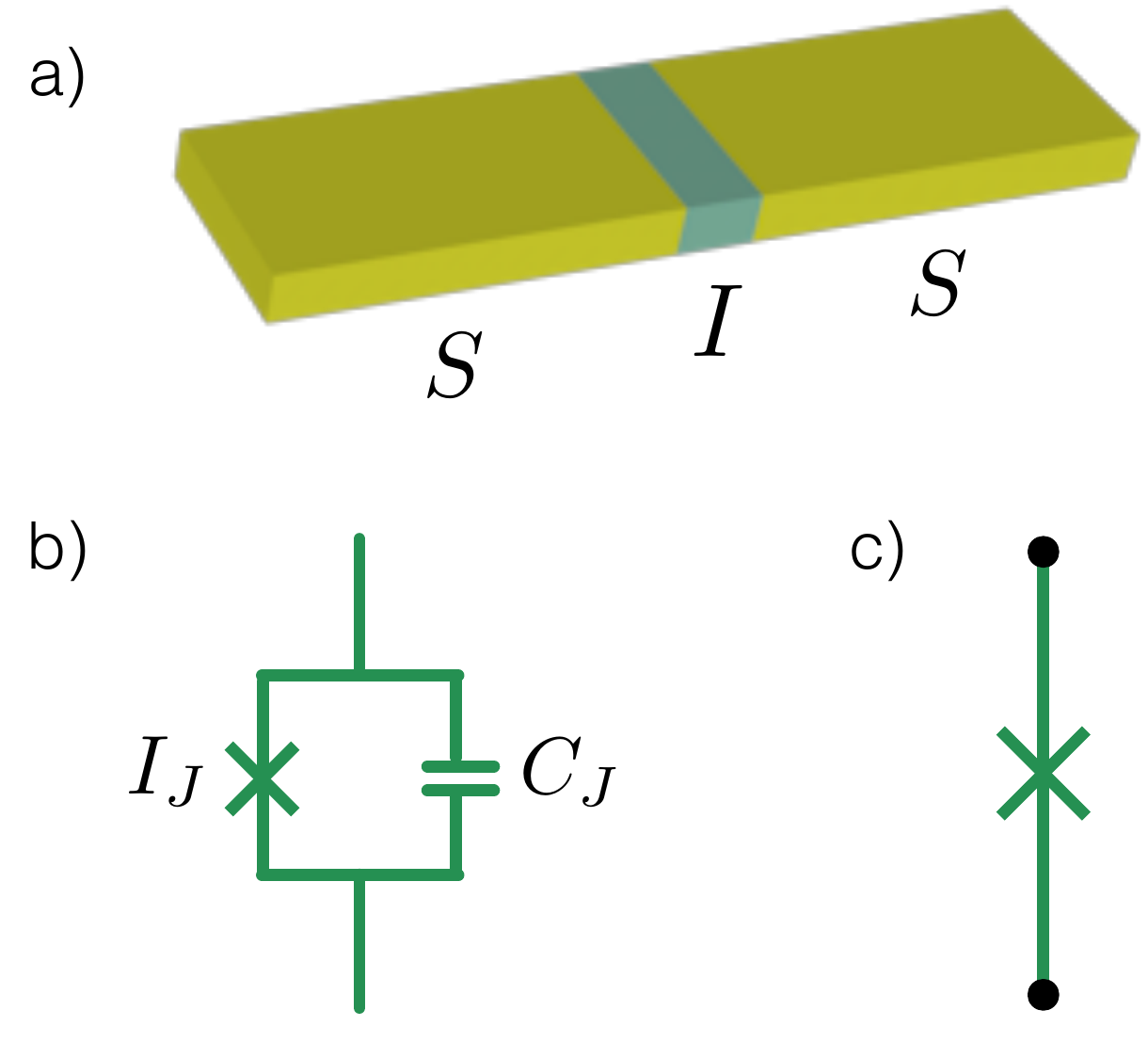}
\caption{a) Schematic representation of a Josephson junction. It consists of two bulk superconductors linked by a thin insulator layer ($\approx 2$\,nm).\,\,\,\,\,\,  b) In the zero-voltage state, a JJ can be characterized by a critical current $I_J$ and a capacitance $C_J$. c) In circuit network theory, a JJ can be described by a single cross that links two nodes.}
\label{fig:1} 
\end{figure}

\begin{enumerate}[{(i)}]
\item \textit{Ultra-low dissipation}. In an integrated circuit, all the metallic parts have to be made out of superconducting materials with negligible resistance at the qubit operating temperature and at the qubit frequency transition. In particular, current experiments make use of low temperature superconductors \cite{Tinkham} such as aluminum or niobium which in turn allow quantum signals to propagate without experiencing dissipation, thus any encoded quantum information may preserve its coherence.      \\
\item \textit{Ultra-low noise}. In a physical realization one can access qubit energies $\hbar\omega_q$ that belong to the range $1-10$\,GHz. In order to avoid thermal fluctuations that may spoil the quantum coherence of qubits, integrated circuits must be cooled down to temperatures of about $T\approx20$\,mK. In general, the energy scales that appear in the system should satisfy the conditions $kT\ll\hbar\omega_q$ and $\hbar\omega_q\ll\varDelta$, where $\varDelta$ is the energy gap of the superconducting material.  \\
\item \textit{Nonlinear, non dissipative elements}. In order to engineer and manipulate two-level systems, it is necessary the access to a device that allows unequal spaced energy levels at the zero-voltage state, where no dissipative current flows through it. These conditions are matched by tunnel junctions \cite{OrlandoBook} as depicted in Fig.~\ref{fig:1}. The Josephson junction (JJ) is a device that consists of two bulk superconductors linked by a thin insulator layer, typically of $1-2$\,nm.         
\end{enumerate}   

\subsection{Hamiltonian Description of a Circuit Network}
\label{subsec:2}

\begin{figure}[t]
\sidecaption
\includegraphics[scale=.6]{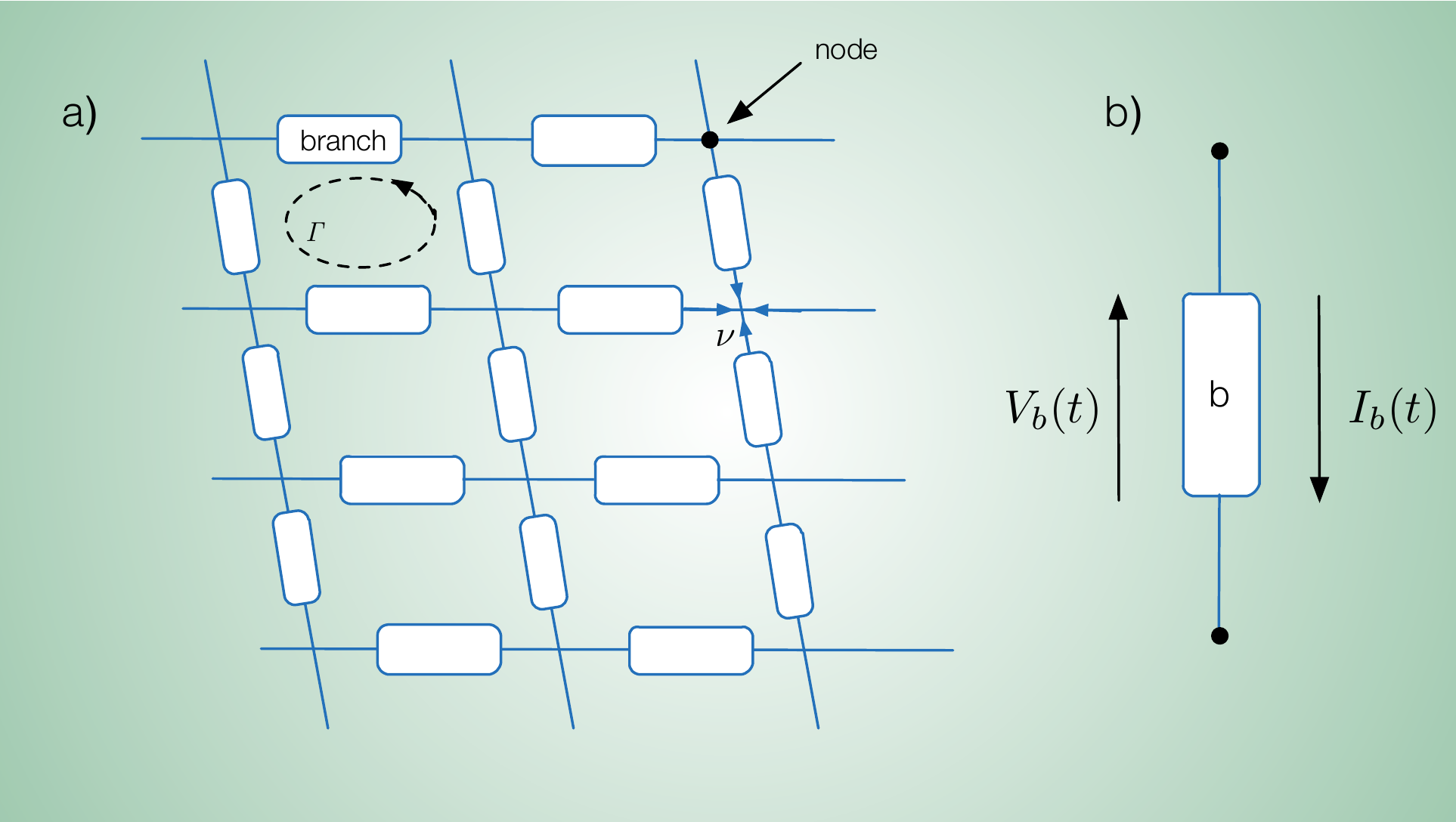}
\caption{a) Electrical circuit network formed by an array of branches and nodes. Each branch may contain an inductor, a capacitor, or a Josephson junction. b) A single branch is characterized by a voltage drop $V_b(t)$ and a current $I_b(t)$. As in a classical electrical circuit, one has to choose the sign convention for currents and voltages, and the Kirchhoff's laws allow us to compute the classical motion equations.}
\label{fig:2}   
\end{figure}

In general, an electrical circuit network is formed by an array of branches and nodes as depicted in Fig.~\ref{fig:2}a. Each branch may contain linear or non linear devices such as inductors, capacitors, or Josephson junctions, and it can be characterized by a branch flux ($\phi_b$) and a branch charge ($q_b$) which are defined in terms of the branch voltages and branch currents (see Fig.~\ref{fig:2}b) by
\begin{eqnarray}
\phi_b(t) &=& \int^{t}_{t_0} dt'\,V_b(t'),\\
q_b(t) &=& \int^{t}_{t_0} dt'\,I_b(t').
\end{eqnarray}

However, these variables do not constitute the degrees of freedom of the circuit because they are linked by the circuit topology through Kirchhoff's laws. Indeed, the sum of all voltages around a closed path $\Gamma$ has to be zero $\sum_{\Gamma[b]} V_{b} =0$. In addition, the sum of all currents of branches tied to a node has to be zero $\sum_{{\nu}[b]} I_{b} =0$, where $\nu$ determines a specific node of the network. 

In order to describe the dynamics of an electrical circuit, one should identify the independent coordinates $\phi_n$, their associated velocities $\dot{\phi}_n$, and to formulate the corresponding Lagrangian $L(\phi_n,\dot{\phi}_n)=T-V$, where $T$ and $V$ stand for the kinetic and potential energies, respectively. A detailed analysis of this procedure can be found in Refs. \cite{DLesHouches,DiVincenzo2004,Girvin2012}, and here we summarize the main concepts. 

\begin{figure}[t]
\sidecaption
\includegraphics[scale=.75]{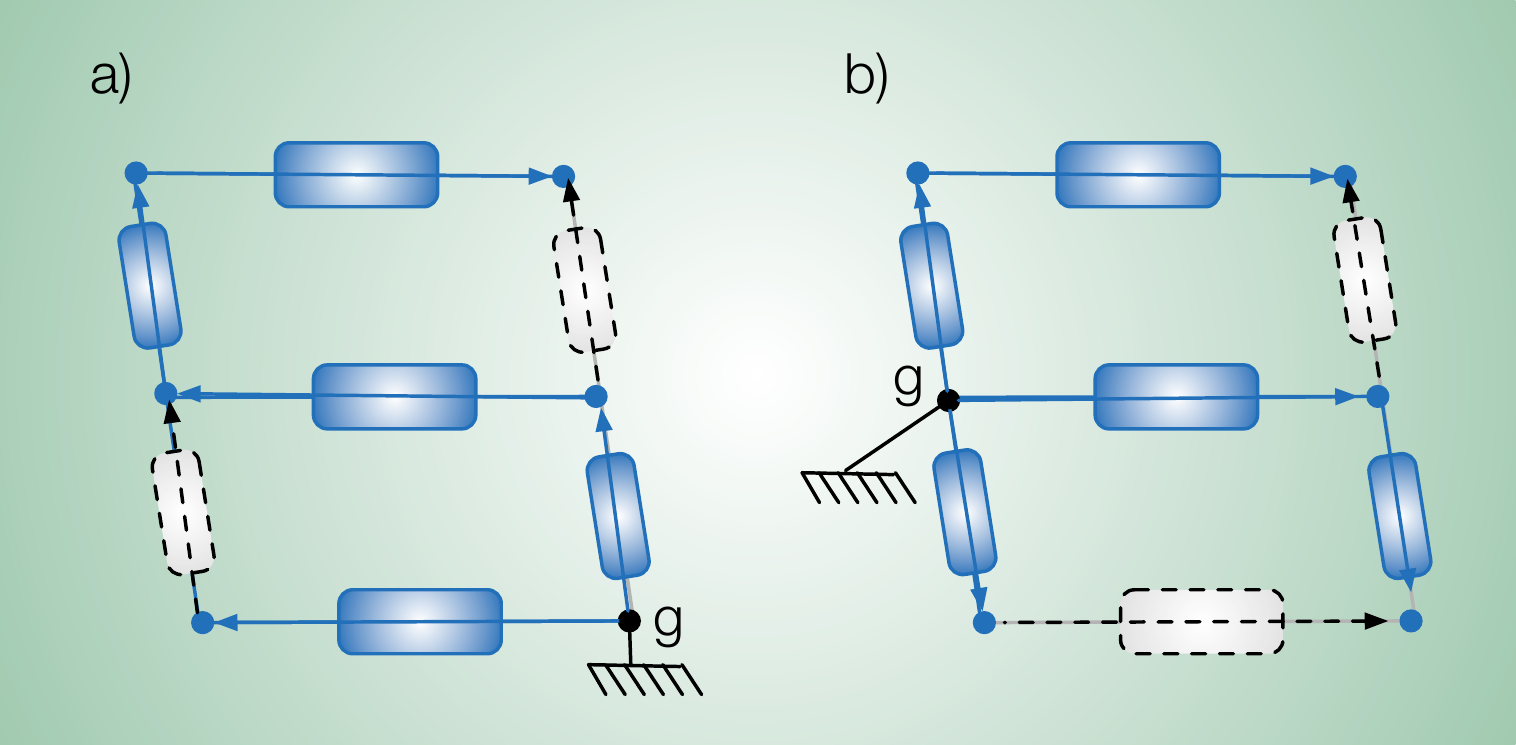}
\caption{Two possible choices of the spanning tree in an electrical circuit. From the ground node $\phi_N=0$, one chooses single paths that pass through a branch (solid branches) to connect actives nodes. The remaining branches (dashed branches) that close a loop have a special treatment due to the flux quantization condition.}
\label{fig:3}
\end{figure}

In circuit theory, one can define \textit{node fluxes} ($\phi_n$) which are variables located at the nodes of the network. These variables depend on a particular description of the topology of the circuit, and such a description is based on the \textit{spanning tree} concept presented in Ref. \cite{DLesHouches}. Specifically, one of the nodes is chosen to be the reference by letting them to act as the ground, say $\phi_N=0$. From the ground node one chooses a unique path that connects to the \textit{active nodes} without closing loops. Figure~\ref{fig:3} shows two possible choices of \textit{node fluxes} and the \textit{spanning tree} (\textit{T}). The latter defines two kind of branches in the network: the set of branches that belong to the spanning tree (blue branches), and the set of closure (\textit{C}) branches each associated with an irreducible loop (grey branches). In terms of node fluxes, each branch belonging to the spanning tree can be defined as $\phi_{b(T)}=\phi_{n+1}-\phi_n$. In addition, the closure branches must be treated in a special way because of the constraint imposed by the flux quantization \cite{Tinkham}. The above condition establishes that for a closed loop threaded by an external magnetic flux $\varPhi_{\rm{ext}}$, the sum of all flux branches that belong to that loop satisfies the rule $\sum_{\Gamma[b]} \phi_{b} - \varPhi_{\rm{ext}} =m\varPhi_0$, where $\varPhi_0=h/2e$ is the flux quantum, and $m$ is an integer. In this sense, the treatment for a closure branch follows the relation $\phi_{b(C)}=(\phi_{n+1}-\phi_n)-\varPhi_{\rm{ext}}$.
   
The building blocks of a general electrical circuit are capacitors, inductors, and Josephson junctions. From these elements it is possible to formulate the corresponding Lagrangian by taking into account their corresponding energies as follows

\begin{eqnarray}
T &=& \frac{C_j}{2}(\dot{\phi}_{n+1}-\dot{\phi}_{n})^2,\\
V &=& \frac{1}{2L_j}(\phi_{n+1}-\phi_{n})^2,\\
V_{JJ} &=&\frac{C_{j,J}}{2}(\dot{\phi}_{n+1}-\dot{\phi}_{n})^2-E_{j,J}\cos\Big(\frac{\phi_{n+1}-\phi_{n}}{\varphi_0}\Big),
\end{eqnarray} 
where $C_j$, $L_j$, $C_{j,J}$, and $E_{j,J}$ stand for the $j$th capacitance, inductance, Josephson capacitance, and Josephson energy, respectively. This procedure allows us to know the matrix capacitance of the system which, in turn, allows us to formulate the Hamiltonian through a Legendre transformation. 

\section{Circuit Quantum Electrodynamics}
\label{sec:circuitQED}
Nowadays, quantum technologies offer a testbed for fundamentals and novel applications of quantum mechanics. In particular, circuit QED \cite{Blais2004,Wallraff2004,Chiorescu2004} has become a leading platform due to its controllability and scalability, with different realizations involving the interaction between on-chip microwave resonators and superconducting circuits which in turn allow to implement transmon, flux, or phase qubits \cite{DSScience}. Here, we briefly describe the interaction between a coplanar waveguide resonator (CWR) and a superconducting circuit be transmon \cite{Koch2007} or flux qubit \cite{Mazo1999}. This in turn will permit us to introduce the \textit{Jaynes-Cummings (JC) Hamiltonian} \cite{JC1963} and the \textit{quantum Rabi Hamiltonian} \cite{Braak2011} that form the building blocks for analog quantum simulations of many-body physics. In addition, we also describe some physical mechanism for the interaction between microwave resonators, allowing us to consider circuit QED lattices.      

\begin{figure}[t]
\sidecaption
\includegraphics[scale=.33]{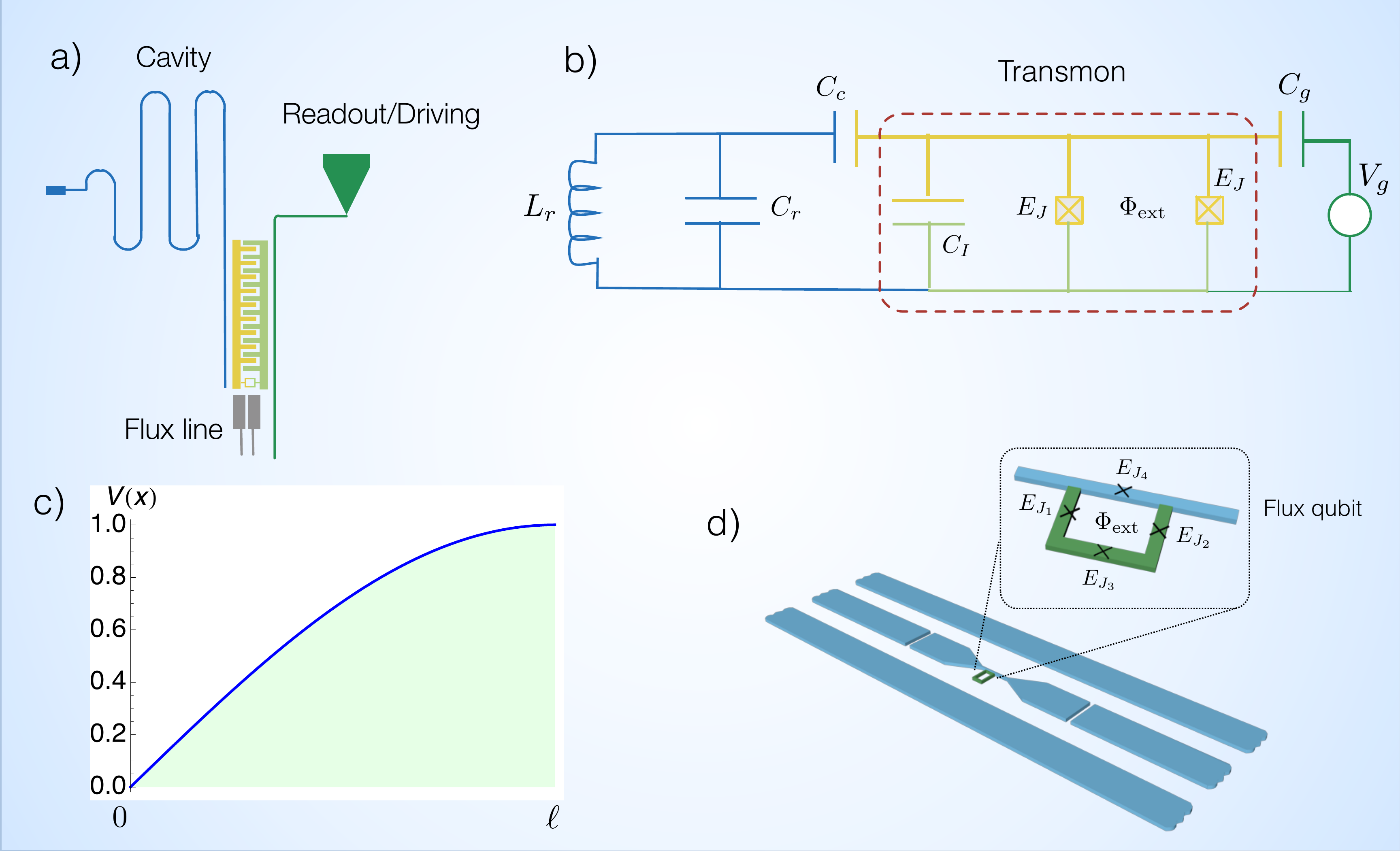}
\caption{Schematic of circuit QED with a transmon qubit. a) A $\lambda/4$ cavity is capacitively coupled to a transmon device. It is also shown the flux line that provides an external magnetic field threading a SQUID loop, and an additional cavity for measuring and driving the qubit. b) Effective lumped circuit element of the cavity-transmon system. c) Voltage distribution for a $\lambda/4$ cavity, where $\ell$ stands for the cavity length. d) A flux qubit formed by three Josephson junctions is galvanically coupled to an inhomogeneous cavity by means of a fourth junction with Josephson energy $E_{J_4}$. This kind of coupling allows us to reach the ultrastrong coupling regime.}
\label{fig:4} 
\end{figure}

\subsection{Circuit QED with a \textit{Transmon} Qubit}
\label{subsec:2.1}

Circuit QED with a \textit{transmon} qubit finds applications in quantum information processing (QIP), where it is possible to implement single- and two-qubit quantum gates~\cite{Majer2007,Leek2009,Blais2007,DiCarlo2009,Haack2010,Bialczak2010,Yamamoto2010}, and three-qubit entanglement generation~\cite{Fink2009,Neely2010,DiCarlo2010}. These proposals find a common basis in the light-matter interaction described by the Jaynes-Cummings interaction \cite{JC1963}.

Figure \ref{fig:4}a represents the electrostatic coupling between a quarter-wave cavity and a transmon qubit, as implemented in current experiments \cite{Groen2013}. For the lowest cavity eigenfrequency, the spatial distribution of the voltage $V(x)$ follows the profile shown in Fig.~\ref{fig:4}c. In this sense, the transmon has to be located at the cavity end to assure a maximum coupling strength. Figure~\ref{fig:4}b shows the effective circuit for the above situation. Two Josephson junctions with capacitance $C_J$ and Josephson energy $E_J$ are shunted by an additional large capacitance $C_I$. This system is coupled to the cavity, with capacitance and inductance $C_r$ and $L_r$, by a comparably large gate capacitance $C_c$. In addition, the transmon is coupled to an external source of dc voltage $V_g$ through the capacitance $C_g$. It is noteworthy to mention that the capacitances $C_I$, $C_c$, and $C_g$ represent effective quantities seen by the transmon, see Ref. \cite{Koch2007}.

The effective Hamiltonian of the joint cavity-transmon system reads
\begin{equation}
H = 4E_C(n-n_g)^2 - E_J \cos\varphi +\hbar\omega_r a^{\dag}a + 2i\beta eV^0_{\rm rms}n(a-a^{\dag}),
\end{equation}
where $n$ and $\varphi$ stand for the number of Cooper pairs transferred between the islands and the gauge-invariant phase difference between them, respectively. In addition, $a(a^\dag)$ annihilates(creates) a single photon of frequency $\omega_r$, $E_C=e^2/2C_\Sigma$ is the charging energy, with $C_\Sigma=C_I+C_J+C_c+C_g$ is the total capacitance associated with the transmon, $n_g=C_gV_g/2e$ is the effective offset charge, and $V^{0}_{\rm rms}=\sqrt{\hbar\omega_r/2C_r}$. 
  
The transmon qubit is less sensitive to charge noise due to an added shunting capacitance $C_I$ between the superconducting islands \cite{Koch2007}. This lowers $E_C$, resulting in an energy ratio of $E_J/E_C\approx 50$. In this case, the two lowest energy levels have an energy splitting that can be approximated by

\begin{equation}
\hbar\omega_{q}\approx\sqrt{8E_CE^{\rm max}_J|\cos(\pi\varPhi_{\rm ext}/\varPhi_0)|}-E_C.
\end{equation}  
Notice that the qubit energy can be tuned by an external magnetic flux $\varPhi_{\rm ext}$ applied to the superconducting quantum interference device (SQUID), see Fig.~\ref{fig:4}b. In this two-level approximation, the dynamics of the cavity-transmon system can be described by the Jaynes-Cummings Hamiltonian \cite{JC1963}

\begin{equation}     
H_{\rm JC} = \frac{\hbar\omega_q}{2}\sigma_z + \hbar\omega_ra^{\dag}a+\hbar g(\sigma^{+}a + \sigma^{-}a^{\dag}),
\label{JCHamiltonian} 
\end{equation}
which exhibits a continuous $U(1)$ symmetry. Here, the two-level system is described by the Pauli matrices $\sigma_j$ $(j=x,y,z)$, and $g$ is the cavity-qubit coupling strength. Notice that working at or near resonance $(\omega_q\sim\omega_r)$, the above Hamiltonian holds true in the rotating-wave approximation (RWA) where the system parameters satisfy the condition $\{g,|\omega_q-\omega_r|\}\ll\omega_q+\omega_r$.   
    
\subsection{Circuit QED with a Flux Qubit}
\label{subsec:2.2}
Circuit QED can also be implemented by means of the inductive interaction between the persistent-current qubit  \cite{Mazo1999}, or flux qubit, and a microwave cavity \cite{Chiorescu2004}. This specific scenario has pushed the technology to reach the \textit{ultrastrong coupling} (USC) regime \cite{Ciuti2005,Bourassa2009,Ciuti2010a,Ciuti2010b,Niemczyk2010,Pol2010} of light-matter interaction, where the qubit-cavity coupling strength reaches a considerable fraction of the cavity frequency. In particular, two experiments have shown a cavity-qubit coupling strength $g=0.12\omega_r$ with a flux qubit galvanically coupled to an inhomogeneous cavity \cite{Niemczyk2010} (see Fig.~\ref{fig:4}d), and to a lumped circuit element \cite{Pol2010} where the Bloch-Siegert shift has been observed. In both experiments, the RWA does not allow to explain the observed spectra. However, it has been shown that the system properties can be described by the quantum Rabi Hamiltonian \cite{Braak2011}

\begin{equation}
H_{\rm Rabi} = \frac{\hbar\omega_q}{2}\sigma_z + \hbar\omega_ra^{\dag}a+\hbar g\sigma_x(a + a^{\dag}).
\label{RabiHamiltonian}
\end{equation}

Unlike the JC dynamics in Eq.~(\ref{JCHamiltonian}), the quantum Rabi Hamiltonian exhibits a discrete parity ($Z_2$) symmetry which establishes a paradigm in the way of understanding the light-matter interaction. For instance, it has been shown in Ref. \cite{Beaudoin2011} that the dissipative dynamics is not longer described by standard master equations of quantum optics \cite{OpenQM}. In addition, the USC regime may have applications such as parity-protected quantum computing \cite{Nataf2011}, and ultrafast quantum gates \cite{Romero2012}. 

The quantum Rabi Hamiltonian allows also to describe the \textit{deep strong coupling} (DSC) regime \cite{Casanova2010}, where the coupling strength is similar or larger than the cavity frequency. This coupling regime has interesting consequences in the breakdown of the Purcell effect \cite{DeLiberato2014}, and it has also been simulated in a waveguide array \cite{Crespi2012}.       
      
\subsection{Cavity-Cavity Interaction Mechanisms}
\label{subsec:2.3}
\begin{figure}[h]
\sidecaption[t]
\includegraphics[scale=.4]{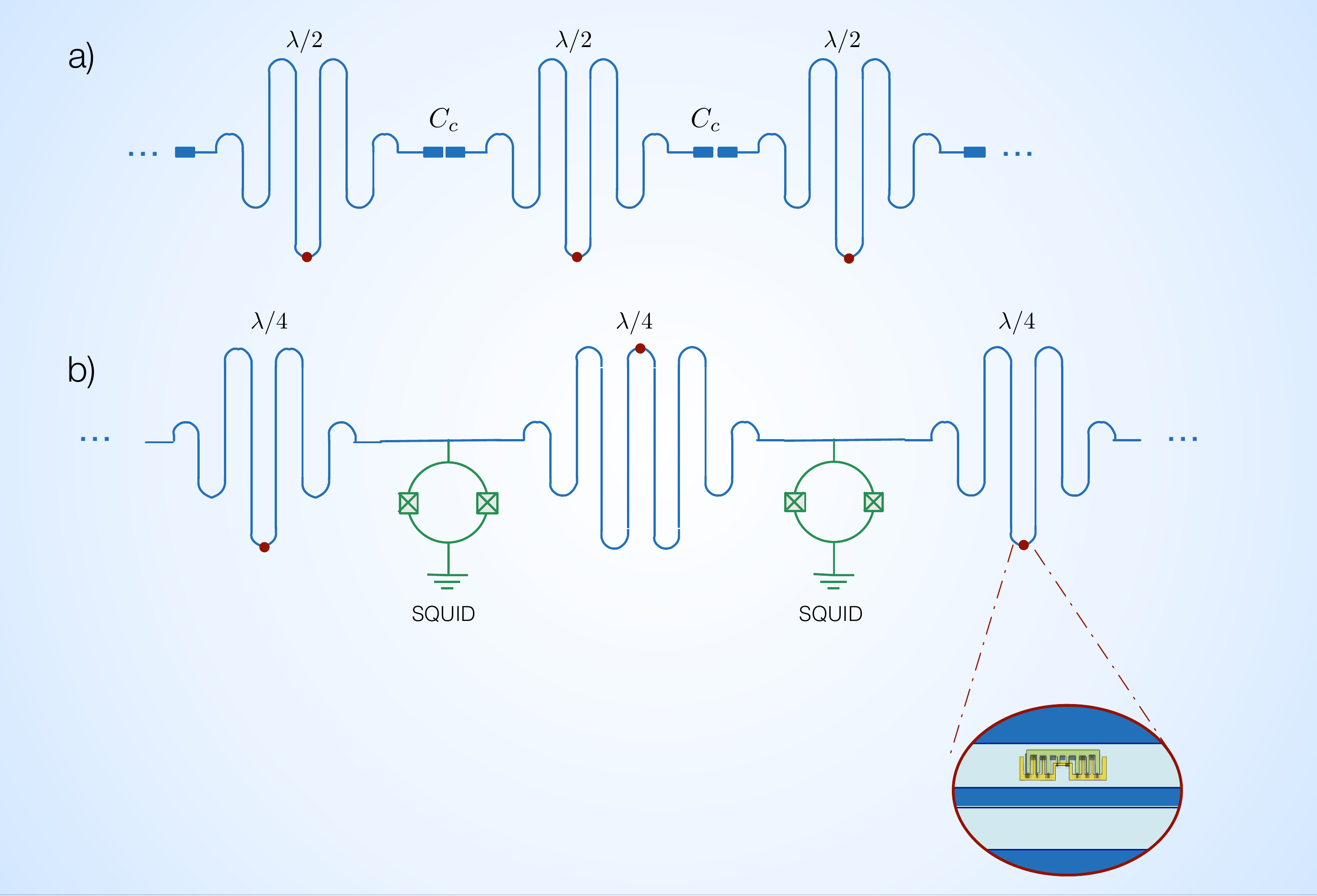}
\caption{Schematic of cavity-cavity interaction. a) Capacitive coupling of half wave cavities. b) Current-current coupling of quarter wave cavities. The latter is mediated by a SQUID device. Inset: red dots represent a transmon qubit capacitively coupled to a microwave cavity.}
\label{fig:5} 
\end{figure}

The versatility of superconducting circuits allow us to design complex arrays involving the interaction among several microwave cavities. There are two possible physical mechanisms to couple them, that is, by means of the capacitive coupling of two half wave cavities (Fig.~\ref{fig:5}a), or two quarter wave cavities via current-current coupling mediated by a SQUID (Fig.~\ref{fig:5}b). In both cases, it is possible to show that the cavity-cavity interaction can be described by a nearest-neighbor Hamiltonian

\begin{equation}
H=\hbar\sum_{n}\omega_{n} a^\dag_{n} a_{n}+\hbar\sum_{\langle n,n'\rangle}J_{nn'} a^\dag_{n} a_{n'},
\label{hopping}
\end{equation}     
where $a_n(a^{\dag}_n)$ stands for the annihilation(creation) operator associated with the $n$th mode of frequency $\omega_n$. Here, we assume the RWA in the hopping term provided by the condition $\{J_{nn'},|\omega_n-\omega_{n'}|\}\ll\omega_n+\omega_{n'}$, and the single-mode approximation in each cavity. These are valid assumptions for realistic parameters that determine the hopping amplitude. For instance, if we consider the coupling capacitance $C_c\ll C_r$, where $C_r$ is the total cavity capacitance, it can be shown that \cite{Nunnenkamp2011} 

\begin{eqnarray}
J_{nn'}&=&\frac{1}{2}\sqrt{\omega_n\omega_{n'}}C_c u_{n}(x)u_{n'}(x')\bigg|_{\rm{ends}},
\end{eqnarray}    
where $u_n(x)$ stands for the spatial dependence of the charge distribution along the cavity. 

In the case of Fig.~\ref{fig:5}b, where the coupling is mediated by the SQUID, it has been shown that the hopping amplitude reads \cite{Felicetti2014}
\begin{eqnarray}
J_{nn'}&\propto&\frac{1}{2}\frac{L_J(\varPhi_{\rm ext})}{\sqrt{\omega_n\omega_{n'}}}\partial_xu_{n}(x)\partial_xu_{n'}(x')\bigg|_{\rm{ends}},
\end{eqnarray}
where $L_J(\varPhi_{\rm ext})$ is the flux-dependent Josephson inductance associated with the SQUID. In analogy to the capacitive coupling case, it is assumed the weak coupling regime $L^0_J\ll L_r$ where $L^0_J=\varphi_0^2/E_J$ is the bare Josephson inductance of the SQUID, $L_r$ is the total inductance of the cavity, and $\varphi_0=\varPhi_0/2\pi$ is the reduced flux quantum. It is noteworthy to mention that this kind of coupling mechanism also generates single-mode squeezing in each cavity that may spoil the implementation of Eq.~(\ref{hopping}). This can be avoided by considering an array of cavities with different lengths as depicted in Fig.~\ref{fig:5}b, and by tuning the system parameters to fulfill the RWA \cite{Felicetti2014}. 

\section{Analog Quantum Simulations with Superconducting Circuits}
\label{sec:AnalogQS}
\subsection{Quantum Simulations: the Jaynes-Cummings Regime}
\label{subsec:3.1}
In the previous chapters, we have shown the basic elements of superconducting circuits and circuit QED. In particular, we have introduced some coupling mechanisms that allow us to model a real physical system in terms of the Jaynes-Cummings~(\ref{JCHamiltonian}) Hamiltonian, but also the coupling mechanisms between microwave cavities. Since the microwave technology shows unprecedented scalability, control, and tunability of physical parameters, circuit QED represents a prime candidate to study many-body states of light \cite{CarusottoCiutiReview} through the analog quantum simulation of the Bose-Hubbard (BH) model \cite{Fisher1989,Hartmann2006,Hartmann2007}, the Jaynes-Cummings-Hubbard (JCH) model \cite{Greentree2006,Angelakis2007}, the fractional quantum Hall effect through the implementation of synthetic gauge fields \cite{Nunnenkamp2011,Angelakis2008,Koch2010,Hafezi2014,Grusdt2014, White2012}, spin lattice systems \cite{Cho2008,Kay2008}, and the bosonic Kagome lattice \cite{Houck2012,Underwood2012,Hamed2014}. 

In the analog quantum simulation of the JCH model
\begin{equation}
H_{\rm JCH} = \sum^{N}_{j=1}\omega_0\sigma^{+}_{j}\sigma^{-}_{j} + \sum^{N}_{j=1}\omega a_{j}^{\dag}a_{j}+J\sum^{N}_{\langle i j \rangle}(a_{i}a^{\dag}_{j}+ \rm{H. c.}),
\end{equation}
the whole system is composed of elementary cells that may consist of a transmon qubit capacitively coupled to a microwave resonator \cite{SchmidtKoch2013}, where the dynamics is described by the JC Hamiltonian. In addition, the connection between neighboring cells is achieved by the capacitive coupling between half-wave cavities in a linear array, see Fig.~\ref{fig:5}a. Notice that the two-site JCH model has been already implemented in the lab \cite{Raftery2015}. Remark that more complex geometries \cite{Nunnenkamp2011} may be achieved, thus establishing an additional advantage over quantum optics platforms~\cite{HarocheRaymond}.

Unlike standard setups of cavity QED in the optical or microwave regimes, circuit QED allows us to engineer nonlinear interactions between cavities provided by nonlinear elements such as Josephson junctions. In particular, the cavity-cavity coupling mediated by SQUID devices, as depicted in Fig.~\ref{fig:5}b, may represent a prime candidate on the road of simulating the Bose-Hubbard model with attractive interactions \cite{Hartmann2010,Hartmann2013}, but also the full Bose-Hubbard and extended models may also be simulated \cite{Dimitris_Unpublished}. It is noteworthy that the Bose-Hubbard-dimer model has been already implemented in a circuit QED setup \cite{Eichler2014}. The latter proposal and the experiment presented in Ref. \cite{Raftery2015} encourage the theoretical work for the sake of simulating many-body states of light and matter. In addition, the driven-dissipative dynamics of many-body states of light \cite{DGAngelakis2009, DGAngelakis2010, Carusotto2009,MJHartmann2010,Grujic2012,Grujic2013,Schetakis2013}, and applications of polariton physics in quantum information processing \cite{DGAngelakis2008, Kyoseva2010} may also be simulated with \textit{state-of-the-art} circuit QED technologies. The coupling mechanisms appearing in circuit QED allow us to study interesting variants of standard models of condensed matter physics. For instance, photon solid phases have been analyzed in the out of equilibrium dynamics of nonlinear cavity arrays described by the Hamiltonian 
\begin{equation}
H = \sum_{i}[-\delta a^{\dag}_ia_i+\Omega(a_i+a^{\dag}_i)] - J\sum_{\langle i,j\rangle}(a_ia^{\dag}_j+{\rm H.c.}) 
+ U\sum_{i}n_i(n_i-1) + V\sum_{\langle i,j\rangle}n_{i}n_{j},
\end{equation}
which exhibits Bose-Hubbard interaction as well as nearest-neighbor Kerr nonlinearities \cite{Fazio2013}. The latter is a direct consequence of the nonlinearity provided by the Josephson energy in the SQUID loop. 

Circuit QED technologies allow also to study two-dimensional arrays of coupled cavities, as stated in Refs. \cite{Houck2012,Underwood2012}, with the implementation of the bosonic Kagome lattice. This provides room to the application of powerful numerical techniques such as the \textit{projected entangled-pair states} (PEPS) \cite{Hamed2014,Cirac2008,PerezGarcia2010} with the aim of studying the interplay between light and matter interactions, as well as predicting new many-body states of light. It is noteworthy to mention that the study of a quantum simulator for the Kagome lattice may predict new physics that otherwise would not be accessible with classical simulations.     

\subsection{Quantum Simulations: the USC Regime of Light-Matter Interactions}
In the previous subsection, we have shown the ability of superconducting circuits to simulate many-body states of light, where the building block or unit cell corresponds to a cavity interacting with a two-level system in the strong coupling (SC) regime. In this case, the cavity-qubit coupling strength exceeds any decay rate of the system such as photon losses, spontaneous decay, and dephasing of the qubit \cite{OpenQM}. Circuit QED has also reached unprecedented light-matter coupling strength with the implementation of the USC regime \cite{Ciuti2005,Bourassa2009,Ciuti2010a,Ciuti2010b,Niemczyk2010,Pol2010} and potentially the DSC regime \cite{Casanova2010,DeLiberato2014}. In this sense, it would be interesting to exploit these coupling regimes aiming at building new many-body states of light, where the building block consists of a cavity-qubit system described by the quantum Rabi model in Eq.~(\ref{RabiHamiltonian}).

Recently, it has been pointed out the importance of the counter-rotating terms in order to describe many-body effects in the \textit{Rabi-Hubbard} model \cite{Takada2011,Tureci2012,Schmidt2013}
\begin{equation}
H_{\rm RH} = \sum_{i}H^{(i)}_{\rm Rabi}- J\sum_{\langle i,j\rangle}(a_ia^{\dag}_j+{\rm H.c.}).
\end{equation}
This model exhibits a $Z_2$ parity symmetry-breaking quantum criticality, long order-range superfluid order, as well as the break of the conservation of local polariton number at each site. This leads to the absence of Mott lobes in the phase diagram as compared with the Bose-Hubbard model. The extension of the Rabi-Hubbard model to the two-dimensional case may represent an additional example where a quantum simulator could outperform classical simulations.   

The above results make it necessary to introduce a quantum simulator that provides the quantum Rabi model (QRM) in a controllable way. As stated in Ref. \cite{Ballester2012}, this task can be done by making use of a two-tone driving on a two-level system of frequency $\omega_q$, that interacts with a single mode of a microwave cavity of frequency $\omega$. In the rotating-wave approximation, the Hamiltonian describing the above situation reads
\begin{eqnarray}
H&=&  \frac{\hbar \omega_q}{2} \sigma_z +\hbar  \omega a^\dag a -\hbar  g (\sigma^\dag a + \sigma a^\dag)  \nonumber \\  
&-& \hbar \Omega_1  (e^{i \omega_1 t} \sigma + e^{-i \omega_1 t} \sigma^\dag) - \hbar  \Omega_2 (e^{i \omega_2 t} \sigma + e^{-i \omega_2 t} \sigma^\dag).\label{HamilDriv}
\end{eqnarray}
Here, $\Omega_j$ and $\omega_j$ represent the Rabi amplitude and frequency of the $j$th microwave signal, respectively. The simulation of the QRM can be accomplished in a specific rotating frame as follows. First, we write the Hamiltonian (\ref{HamilDriv}) in the reference frame that rotates the frequency $\omega_1$. This leads to
\begin{eqnarray}
H^{R_1} &=&\hbar  \frac{(\omega_q-\omega_1)}{2} \sigma_z +\hbar  (\omega-\omega_1) a^\dag a - \hbar g  \pare{\sigma^\dag a +\sigma a^\dag} \nonumber \\ & &  \hspace{-13mm} - \hbar \Omega_1  \pare{ \sigma + \sigma^\dag } - \hbar  \Omega_2   \pare{ e^{i (\omega_2-\omega_1) t} \sigma + e^{-i (\omega_2-\omega_1) t} \sigma^\dag } .
\end{eqnarray}
Second, we go into the interaction picture with respect to $H_0^{R_1} = - \hbar \Omega_1  \pare{ \sigma + \sigma^\dag} $ such that $H^{I} (t) = e^{i H_{0}^{L_1} t/\hbar } \pare{H^{R_1}   - H_0^{R_1} }    e^{-i H_{0}^{R_1} t/\hbar } $. The above transformation can be implemented by means of a Ramsey-like pulse as described in Ref. \cite{Ballester2012}. In the dressed-spin basis, $\ket{\pm} = \pare{\ket{g} \pm \ket{e} }/\sqrt2$, the interaction Hamiltonian we can be written as
\begin{eqnarray}
H^{I} (t) &=& -\hbar  \frac{(\omega_q-\omega_1)}{2} \pare{ e^{-i 2  \Omega_1  t} \proj{+}{-}  + {\rm H.c.}}   + \hbar (\omega-\omega_1) a^\dag a \nonumber  \\
  & -& \frac{\hbar g }{2} \left( \left\{  \proj{+}{+} - \proj{-}{-} + e^{-i 2  \Omega_1 t} \proj{+}{-} \right. \right. \nonumber \\
 & & -  \left.\left. e^{i 2  \Omega_1  t}\proj{-}{+}  \right\} a + {\rm H.c.} \right) \nonumber \\  
  &-&  \frac{\hbar \Omega_2 }{2} \left(   \left\{  \proj{+}{+} - \proj{-}{-} - e^{-i 2  \Omega_1  t} \proj{+}{-}  \right. \right. \nonumber \\ & & \left. \left. + e^{i 2 \Omega_1  t}\proj{-}{+}  \right\} e^{i (\omega_2-\omega_1) t} + {\rm H.c.} \right). \label{HI1}
\end{eqnarray}

\begin{figure}[t]
\includegraphics[width=0.5\textwidth]{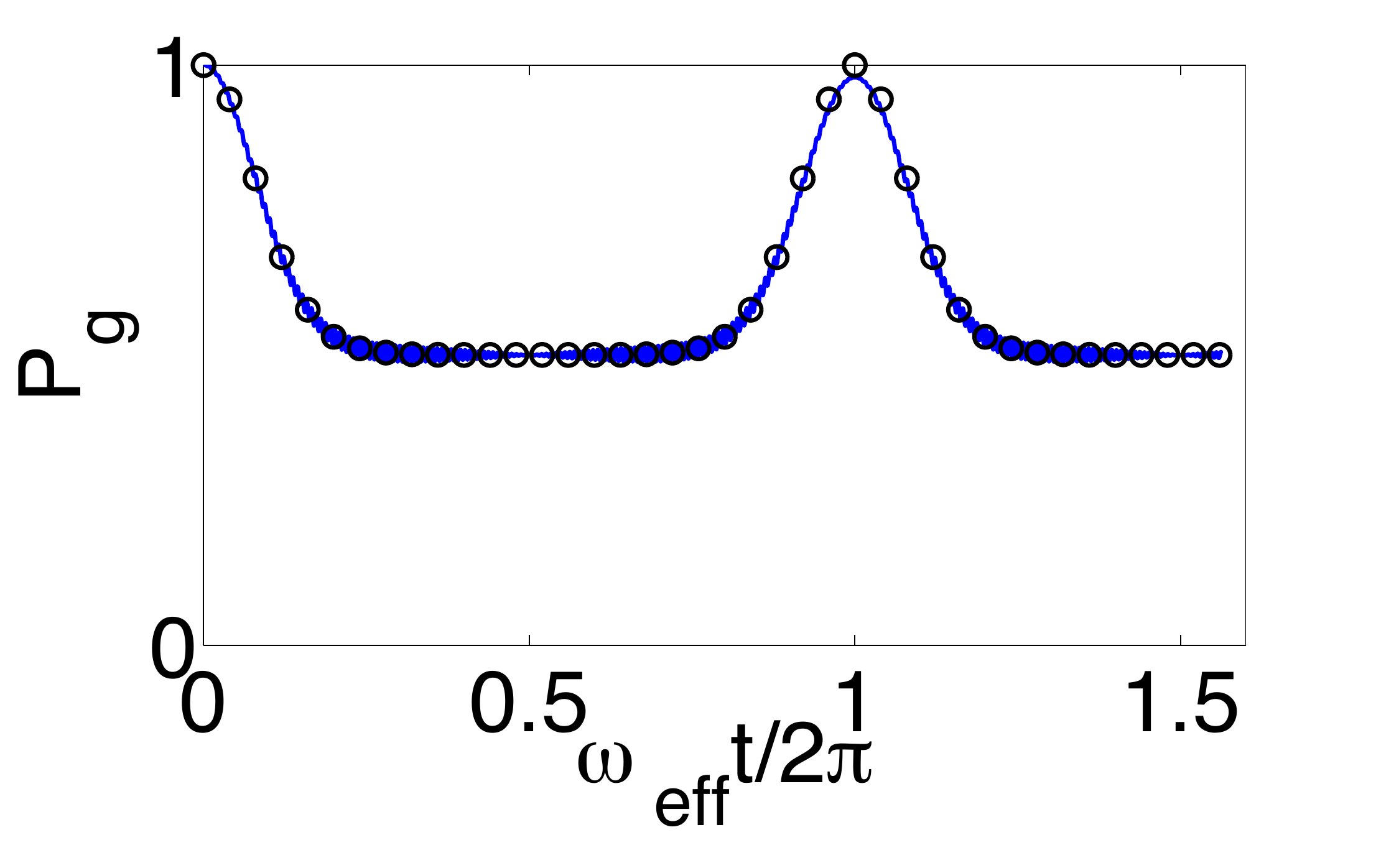}
\includegraphics[width=0.5\textwidth]{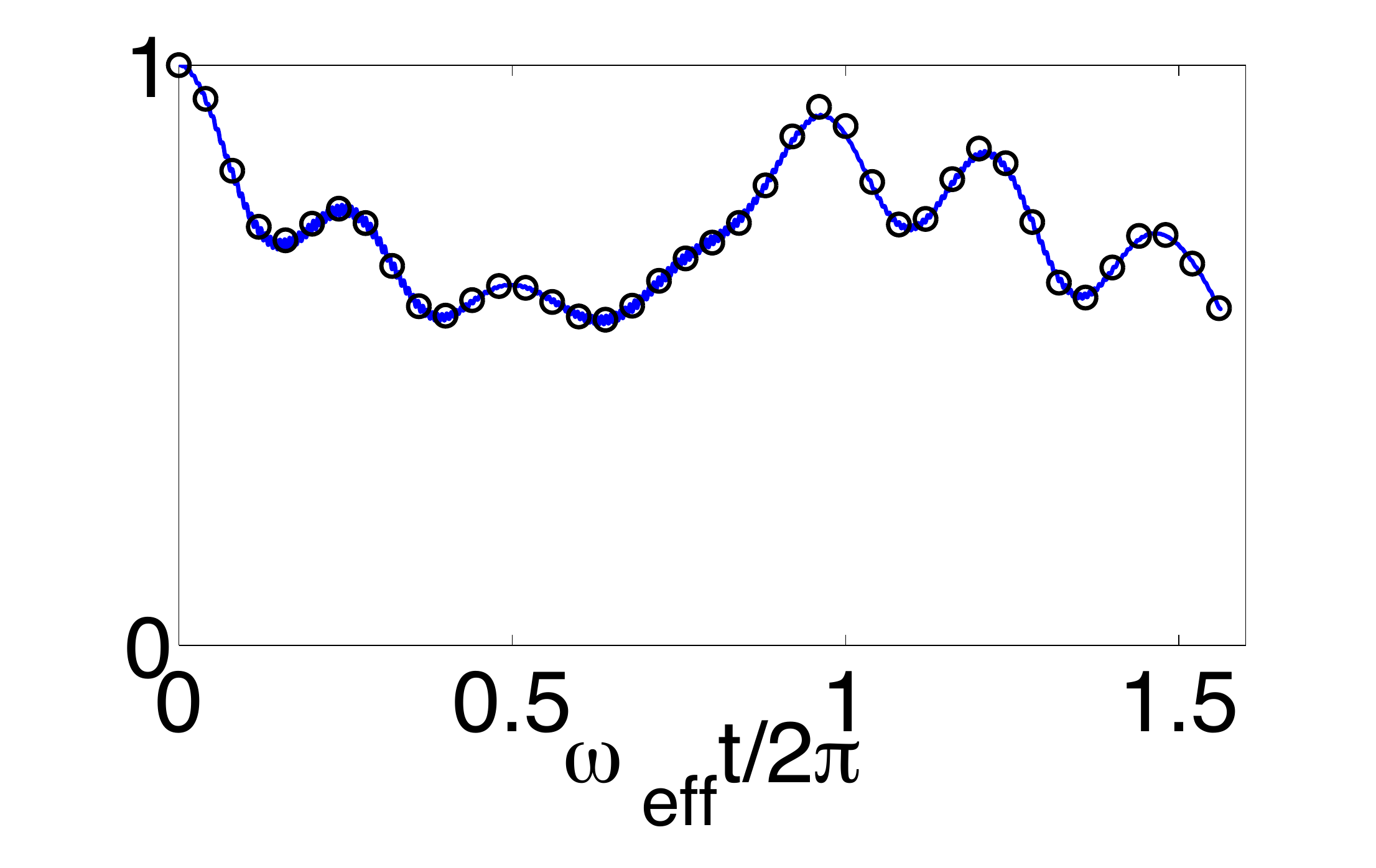}
\caption{\label{cqed:figs}
The population of the ground state for the two-level system, $P_{g}(t)$, has been calculated by integrating the exact (solid line) dynamics in Eq.~(\ref{HamilDriv}) and the effective (circles) Hamiltonian dynamics in Eq.~(\ref{HamilEff}). We have considered two different cases: {\bf (left panel)} $\Omega_2  =0$;  {\bf (right panel)} $\Omega_2  =2\pi\times10\,$MHz. The simulated interaction strength corresponds to $g_{\rm eff} / \omega_{\rm eff} =1$. Figure from Ref. \cite{Ballester2012}, used under the terms of the Creative Commons Attribution 3.0 licence.} 
\end{figure}

Third, if we tune the external driving frequencies as $\omega_1-\omega_2=2 \Omega_1$, and for a strong first driving, $ \Omega_1$, the Hamiltonian (\ref{HI1}) can be represented as
\begin{eqnarray}
H_{\rm eff}
=  \hbar (\omega-\omega_1) a^\dag a + \frac{\hbar\Omega_2}{2} \sigma_z   -  \frac{\hbar g}{2} \sigma_x \pare{a+a^\dag},  \label{HamilEff}
\end{eqnarray}
which corresponds to the quantum Rabi model. For values $\Omega_2  \sim (\omega-\omega_1) \sim g/2$, the original cavity-qubit system is capable of simulating the dynamics associated with the USC/DSC regime. The simulated interaction strength corresponds to the ratio $g_{\rm eff}/\omega_{\rm eff}$, where $g_{\rm eff} \equiv g/2$ and $ \omega_{\rm eff}\equiv\omega-\omega_1$. Figure~\ref{cqed:figs} shows the two-level system dynamics for an effective ratio $g_{\rm eff}/\omega_{\rm eff}=1$ and two different values of the effective qubit frequency $\Omega_2$. The initial condition is the ground state of the qubit and the vacuum state for the field, $|\psi(0)\rangle=|g\rangle\otimes|0\rangle$. This dynamics corresponds to the one predicted in Ref. \cite{Casanova2010} for the deep strong coupling regime.

This quantum simulation may pave the way for building a complete toolbox of complex cavity arrays where the unit cell can be tuned, at will, from the strong coupling regime, described by the Jaynes-Cummings model, to the USD/DSC regime described by the quantum Rabi model.

\subsection{Quantum Simulations of Quantum Relativistic Mechanics}

The quantum simulation of the quantum Rabi model \cite{Ballester2012} allows us to access a wide range of physical phenomena such as cat-state generation and simulating relativistic quantum mechanics on a chip \cite{Pedernales2013}. The latter can be achieved in a similar way as the QRM by applying three classical microwaves, that is, a two-tone driving on a two-level system interacting with a single cavity mode in the SC regime, and a driving on the cavity mode. This process is modeled by the Hamiltonian
\begin{eqnarray}
H &=& \frac{\hbar\omega_q}{2}\sigma_z + \hbar\omega a^{\dag}a - \hbar g(\sigma^{+}a+\sigma^{-}a^{\dag})- \hbar \Omega(\sigma^{+}e^{-i(\omega t + \varphi)}+\sigma^{-}e^{i(\omega t+\varphi)})\nonumber \\ 
&-& \lambda(\sigma^{+}e^{-i(\nu t + \varphi)}+\sigma^{-}e^{i(\nu t+\varphi)})+\hbar\xi(ae^{i\omega t} + a^{\dag}e^{-i\omega t}),
\label{HL1}
\end{eqnarray}
where $\Omega$, $\lambda$, and $\xi$ stand for the driving amplitudes, $\omega$ is the resonator frequency, and $\nu$ is the driving frequency, respectively. As stated in Ref. \cite{Pedernales2013}, if we consider a strong microwave driving $\Omega\gg \{g,\lambda\}$ and the condition $\omega-\nu=2\Omega$, the effective Hamiltonian, in the rotating frame, reads 
\begin{equation}
H_{\rm eff}=\frac{\hbar\lambda}{2}\sigma_z+\frac{\hbar g}{\sqrt{2}}\sigma_y\hat{p}+\hbar\xi\sqrt{2}\hat{x},
\end{equation}
where $\hat{x}=(a+a^{\dag})/\sqrt{2}$ and $\hat{p}=i(a^{\dag}-a)/\sqrt{2}$ are the field quadratures satisfying the commutation relation $[\hat{x},\hat{p}]=i$. The above Hamiltonian describes a $1+1$ Dirac particle in a linear external potential $U=\hbar\xi\sqrt{2}\hat{x}$, where the terms $\hbar g/\sqrt2$ and $\hbar \lambda / 2$ represent the speed of light and the mass of the particle, respectively.

\begin{figure}[t]
\includegraphics[width=\textwidth]{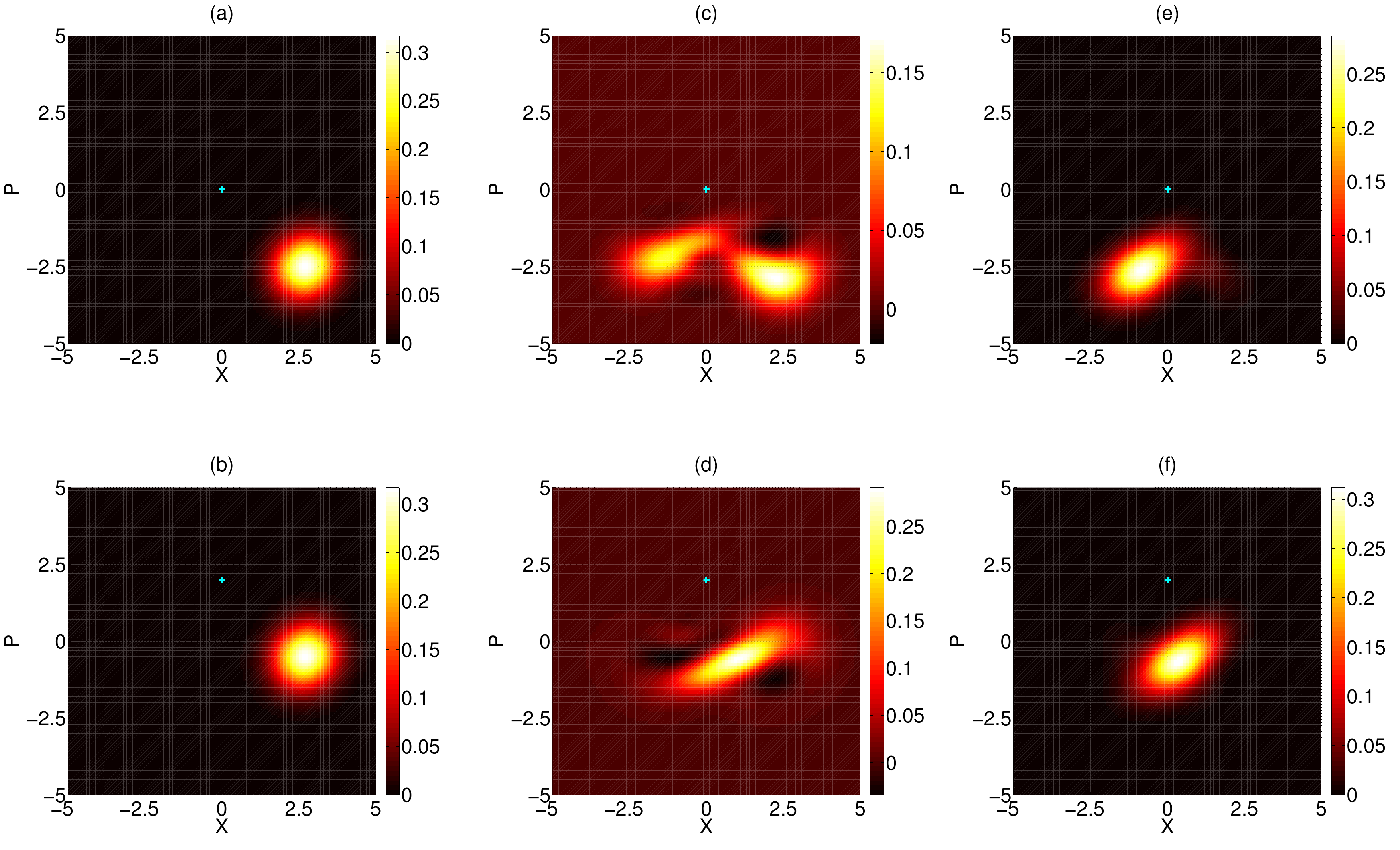}
\caption{\label{klein-figs} Wigner function $W(x,p)$ representation of the field state in the microwave cavity. We have computed the time evolution by means of the the Hamiltonian (\ref{HL1}) for a time of $60$ nsec. We used realistic parameters $g=2\pi\times 10$ MHz, $\Omega=2\pi\times 200$ MHz, $\xi=g/2$. {\bf (a)} $\lambda=0$ with the initial state $\ket{+,0}$; {\bf (b)} $\lambda=0$ with the initial state $\ket{+,\sqrt2i}$; {\bf (c)} $\lambda=\sqrt2 g$ with the initial state $\ket{+,0}$; {\bf (d)} $\lambda=\sqrt2 g$ with the initial state $\ket{+,\sqrt2i}$; {\bf (e)} $\lambda= 4 \sqrt2 g$ with the initial state $\ket{e,0}$; and {\bf (f)} $\lambda= 4 \sqrt2 g$ with the initial state $\ket{e, \sqrt2i}$. Figure from Ref. \cite{Pedernales2013}, used under the terms of the Creative Commons Attribution 3.0 licence.}
\end{figure}

Adding an external potential $U(x)$ allows us to simulate the scattering of a single relativistic particle. In particular, we can start by considering the case of a massless Dirac particle whose Hamiltonian is given by $H_{\rm K} =  \hbar g/\sqrt2  \sigma_y \hat{p} + \hbar\xi\sqrt2 \, \hat{x}$. Figures \ref{klein-figs} (a) and (b) show the evolution of the initial states $\ket{+,0}$ and  $\ket{+,\sqrt2i}$, respectively, where $\ket{+}$ stands for the positive eigenstate of $\sigma_y$ and $\ket{0}$ the vacuum state for the field. Here the field state remains coherent while experiencing two independent displacements along the $\hat{x}-$quadrature proportional to $g/\sqrt2$, and along the $\hat{p}-$quadrature. It is remarkable that the external potential does not modify the rectilinear movement in position representation. This phenomenon corresponds to the \textit{Klein paradox}, which states that a massless Dirac particle may propagate through the potential barrier with probability different from zero. 

As the quantum simulation allows us to tune the physical parameters at will, we can also study the scattering of a massive nonrelativistic Schr\"odinger particle. In this case, the dynamics is governed by the Hamiltonian $H_{\rm NRel} = \hbar \sigma_z \hat{p}^2/\lambda + \hbar\xi \sqrt2 \, \hat{x}  $. Note that this Hamiltonian assures that any initial Gaussian state remains Gaussian as time elapses. This can be seen in Figs. \ref{klein-figs}(e) and (f) where the initial states are $\ket{e,0}$ and $\ket{e,\sqrt2i}$, respectively. The mass of the particle has been chosen such that $\hbar \lambda/2=4\times \hbar g/\sqrt2$. Figure \ref{klein-figs}(e) shows how the particle is scattered backwards by the potential. For the case of Fig. \ref{klein-figs}(f), the particle has an initial positive kinetic energy that allowed it to enter the external potential, though after $60$ nsec it moves backwards.

These two limiting cases have shown a total transmission or reflection. It is natural that a particle with an intermediate mass features only partial transmission/reflection. Figure \ref{klein-figs}(c) shows the scattering of a massive particle with $\hbar \lambda/2=  \hbar g/\sqrt2$ prepared in the initial state $\ket{+,0}$. We see how the wave packet splits into spinor components of different signs which move away from the center. Furthermore, If some initial kinetic energy is provided to the wavepacket, as shown in Fig. \ref{klein-figs}(d) with initial state $\ket{+,\sqrt2i}$, the particle enters the barrier to stop and break up sooner or later.

The simulation of the Dirac Hamiltonian, together with all available technology in circuit QED, may have interesting consequences in the study of many-body states of light. For instance, one may have access to Dirac lattices and their possible extension to Dirac materials \cite{DiracM}. 

\section{Digital Quantum Simulations with Superconducting Circuits}
\label{sec:dig1}
In many situations, the quantum simulator does not evolve according to the dynamics of the system to be simulated. Therefore, it is appropriate to employ digital techniques to emulate a wider variety of quantum systems~\cite{Lloyd96}. Digital quantum simulators are akin to universal quantum computers, with the advantage that in principle with a small number of qubits one will already be able to outperform classical computers. Thus, one does not need to reach thousands of qubits to perform interesting quantum simulations of mesoscopic quantum systems.

The digital quantum simulators are based on the fact that most model Hamiltonians are composed of a finite number of local terms, $H=\sum_{k=1}^N H_k$, where each of them acts upon a reduced Hilbert space, or at least is efficiently implementable with a polynomial number of gates. In these cases, the system dynamics can be obtained via digital decomposition into stroboscopic steps, via Trotter techniques,
\begin{equation}
e^{-iHt}=(e^{-iH_1t/n}...e^{-iH_Nt/n})^n+O(t^2/n).\label{Trotter}
\end{equation}
By making $n$ large, the error can be made in principle as small as desired. Naturally there will be a limit to the size of $n$ that will be given by the finite fidelity of the local gates.   

There have been already a number of experiments on digital quantum simulators, either in quantum photonics~\cite{LanyonQChem}, or in trapped ions~\cite{Lanyon11}. Regarding superconducting circuits, some theoretical proposals for digital quantum simulations have been put forward~\cite{Urtzi14,Laura14}, and we will review these in the next subsections.

\subsection{Digital Quantum Simulations of Spin Systems with Superconducting Circuits}
\label{subsec:dig1}

In this section, we analyze the realization of digital quantum simulations of spin Hamiltonians in a superconducting circuit setup consisting of several transmon qubits coupled to a microwave resonator~\cite{Urtzi14}. Although our protocol is appropriate for every superconducting qubit with sufficiently long coherence time, we consider specifically a transmon qubit device. This kind of qubits are typically used because of its insensitivity to charge fluctuations~\cite{Koch2007}. Nevertheless, depending on the specific phenomena to simulate, one can consider other superconducting qubits for quantum simulations. First, we show how one can simulate the Heisenberg model in a circuit QED setup with state-of-the-art technology. Then, we consider typical simulation times and their associated fidelities with current superconducting qubit technology, showing the potential of superconducting qubits in terms of digital quantum simulators. Finally, we study the necessary resources with realistic parameters for a versatile quantum simulator of spin models able to emulate a general many-body spin dynamics.

Digital methods can be employed to emulate the Heisenberg model with current circuit QED technology.  Even though the latter does not feature the Heisenberg interaction from first principles, one can nevertheless analyze a digital quantum simulation of this model.
We show that a set of $N$ transmon qubits coupled through a resonator is able to simulate Heisenberg interactions of $N$ spins, which in the symmetric-coupling case is given by
\begin{equation}
H=\sum_{i=1}^{N-1}  J\left(\sigma^x_i \sigma^x_{i+1} +\sigma^y_i \sigma^y_{i+1} +\sigma^z_i \sigma^z_{i+1}\right).\label{XYZinhom}
\end{equation}
Here $\sigma^j_i$, $j\in\{x,y,z\}$ are Pauli matrices that refer to the first two levels of the $i$th transmon qubit. 

\begin{figure}
\center
\includegraphics[scale=0.33]{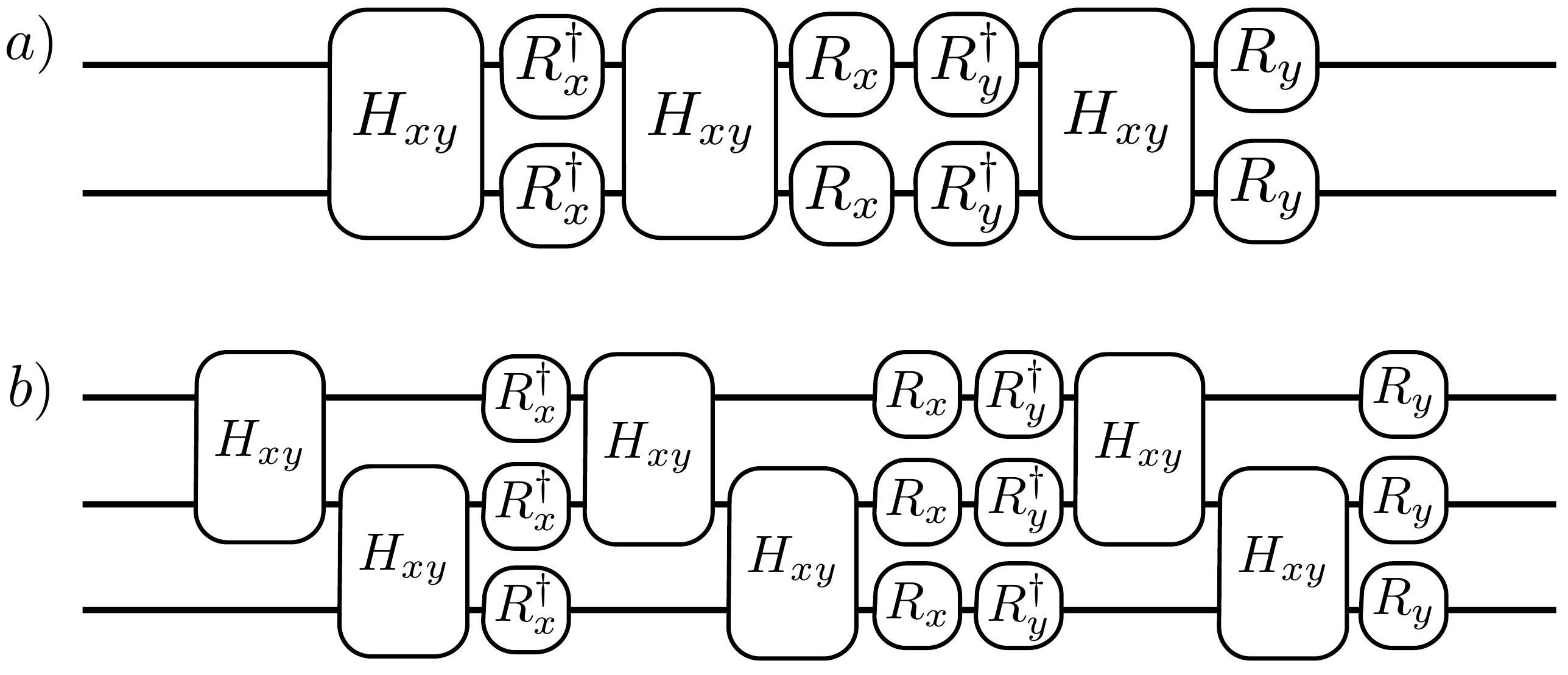}
\caption{Schemes for the proposed digital quantum simulations with superconducting transmon qubits. a) Heisenberg model for two qubits. b) Heisenberg model for three qubits. Here, $R_{x(y)}\equiv R^{x(y)}(\pi/4)$ and $\overline{R}_x\equiv R^x(\pi/2)$. We point out that exchanging each of the $R$ matrices with its adjoint does not affect the protocols. Reprinted with permission from \cite{Urtzi14}, Copyright (2014) American Physical Society. \label{Fig1Heis}} 
\end{figure}

 We start by the simplest case, with only two spins. The XY exchange interaction can be implemented by dispersive coupling of two transmon qubits with a common resonator~\cite{Blais2004,Majer2007,Filipp11}, $H_{12}^{xy}=  J \left( {\sigma_1^+} {\sigma_2 ^-}+  {\sigma_1^-} {\sigma_2^+} \right) =   J/2  \left(\sigma^x_1 \sigma^x_2 +  \sigma^y_1 \sigma^y_2\right)$.
The XY interaction can be mapped through local rotations of the qubits onto the effective Hamiltonians 
\begin{equation}
H_{12}^{xz} = R^x_{12}(\pi/4)H_{12}^{xy}R^{x \dagger}_{12}(\pi /4) = J/2  \left(\sigma^x_1 \sigma^x_2 +  \sigma^z_1 \sigma^z_2 \right),
\end{equation}
\begin{equation}
H_{12}^{yz}=R^{y}_{12}(\pi/4)H_{12}^{xy}R^{y \dagger}_{12}(\pi /4)= J/2  \left(\sigma^y_1 \sigma^y_2 +  \sigma^z_1 \sigma^z_2 \right).
\end{equation} 
Here, $R_{12}^{x(y)}(\pi/4)=\exp[-i\pi/4(\sigma_1^{x(y)}+\sigma_2^{x(y)})]$ is a local rotation of the first and second qubits with respect to the $x(y)$ axis.
The XYZ Heisenberg Hamiltonian $H_{12}^{xyz}$ can thus be performed according to the following protocol (see Fig.~\ref{Fig1Heis}a).
 {\it Step 1.--} The two qubits interact  with the XY Hamiltonian $H_{12}^{xy}$ for a time $t$.
{\it Step 2.--} Single qubit rotations $R^x_{12}(\pi/4)$ are applied to both qubits. 
{\it Step 3.--} The two qubits interact with $H_{12}^{xy}$ Hamiltonian for a time $t$.
{\it Step 4.--} Single qubit rotations $R^{x \dagger}_{12}(\pi/4)$ are applied to both qubits. 
{\it Step 5.--} Single qubit rotations $R^y_{12}(\pi/4)$ are applied to both qubits. 
{\it Step 6.--} The two qubits interact according to the $H_{12}^{xy}$ Hamiltonian for a time $t$.
{\it Step 7.--} Single qubit rotations $R^{y \dagger}_{12}(\pi/4)$ are applied to both qubits. 
Accordingly, the final unitary operator reads
\begin{eqnarray}
U_{12} (t)&=& e^{-i H_{12}^{xy} t} e^{-i H_{12}^{xz} t} e^{-i H_{12}^{yz} t}= e^{-i H_{12}t}.\label{eq8}
\end{eqnarray}
This evolution emulates the dynamics of Eq.~(\ref{XYZinhom}) for two transmon qubits. Furthermore, arbitrarily inhomogeneous couplings can be engineered by performing different evolution times or couplings for the different digital gates. Here we point out that a single Trotter step is needed to obtain a simulation with no digital errors, because of the fact that $H_{12}^{xy}$, $H_{12}^{xz}$, and $H_{12}^{yz}$ operators commute. Accordingly, in this case the only error source will be due to the accumulated gate errors. We consider two-qubit gates with a process fidelity error of about $5\%$ and eight $\pi /4$ single-qubit gates with process fidelity errors of about $1\%$. Therefore, we will have a total process fidelity of this protocol of about $77\%$. Furthermore, the total protocol time for a $\pi/4$ Heisenberg phase is around $0.10$~$\mu$s. Throughout this section, we estimate the protocol times by adding the respective times of all the gates, for which we take into account standard superconducting qubit values.

We now analyze a digital algorithm for the emulation of the Heisenberg dynamics for a system of three spins. In this case, one has to consider noncommuting Hamiltonian gates, involving Trotter errors. This three-spin model can be directly extrapolated to an arbitrary number of spins. We propose the following digital protocol for its realization (see Fig.~\ref{Fig1Heis}b). {\it Step~1.--}~Qubits 1 and 2 couple through XY Hamiltonian for a time $t/l$. {\it Step~2.--}~Qubits 2 and 3 couple through XY Hamiltonian for a time $t/l$. {\it Step~3.--}~The gate $R^x_{i}(\pi/4)$ is applied to each qubit.
{\it Step~4.--}~Qubits 1 and 2 couple through XY Hamiltonian for a time $t/l$.
{\it Step~5.--}~Qubits 2 and 3 couple through XY Hamiltonian for a time $t/l$.
{\it Step~6.--}~The gate $R^{x\dagger}_{i}(\pi/4)$ is applied to each qubit. 
{\it Step~7.--}~The gate $R^y_{i}(\pi/4)$ is applied each qubit. 
{\it Step~8.--}~Qubits 1 and 2 couple through XY Hamiltonian for a time $t/l$.
{\it Step~9.--}~Qubits 2 and 3 couple through XY Hamiltonian for a time $t/l$.
{\it Step~10.--}~The gate $R^{y\dagger}_{i}(\pi/4)$ is applied to each qubit.
Finally, the global unitary evolution operator per Trotter step is given by
\begin{eqnarray}
U_{123}(t/l) =e^{-i H_{12}^{xy} t/l}e^{-i H_{23}^{xy} t/l} e^{-i H_{12}^{xz} t/l}\nonumber e^{-i H_{23}^{xz} t/l} e^{-i H_{12}^{yz} t/l}e^{-i H_{23}^{yz} t/l} \label{eq15}.
\end{eqnarray}
\begin{figure}\center
\includegraphics[scale=0.33]{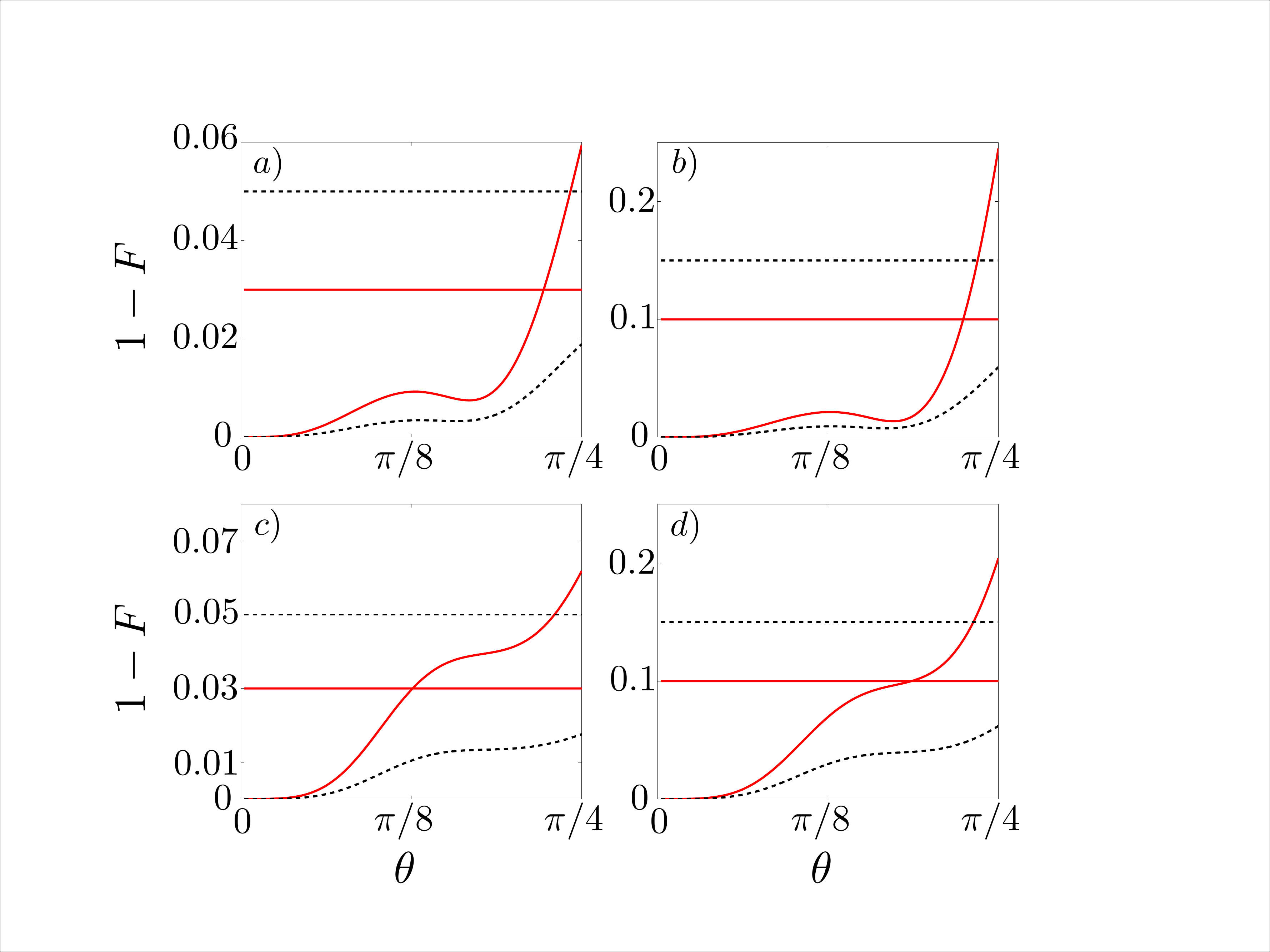}
\caption{Loss of fidelity for the emulated Heisenberg model for three qubits, in the range $\theta=[0,\pi/4]$, $\theta\equiv Jt$. The wavy lines represent digital errors, while the straight lines represent the accumulated gate error due to a step error of $\epsilon$. Solid (dotted) lines are associated with lower (higher) digital approximations $l$. a) $\epsilon=10^{-2}$, $l=3,5$, and b) $\epsilon=5\times10^{-2}$, $l=2,3$. Reprinted with permission from \cite{Urtzi14}, Copyright (2014) American Physical Society.}\label{Fig2Heis}
\end{figure}
Here, the sequence has to be repeated $l$ times following Eq.~(\ref{Trotter}), to implement an approximate dynamics of Eq.~(\ref{XYZinhom}) for the three qubits. Each of these Trotter steps consists of four single-qubit gates at different times (performed collectively upon different sets of qubits) and six XY gates, with a total step time around $0.16$~$\mu$s, well below typical decoherence times in transmon qubits~\cite{Rigetti12}. In Figs.~\ref{Fig2Heis}a and \ref{Fig2Heis}b, we depict the fidelity loss associated with the digital error of the emulated XYZ dynamics for three transmon qubits, together with straight horizontal lines showing the error of the imperfect gates multiplied by the number of Trotter steps, i.e., the total accumulated gate imperfection. One can appreciate time intervals dominated by the digital Trotter error and time intervals where the largest error source in the digital quantum simulation is produced by experimental gate imperfections. One can take into account Hamiltonians with open and periodic boundary conditions, adding an extra coupling between the first and the last spin. Extending our protocol to $N$ transmon qubits with open or periodic boundary conditions, we estimate an upper bound on the second-order digital Trotter error, given by $E_\textrm{open}=24(N-2)(Jt)^2/l$ and $E_\textrm{periodic}=24N(Jt)^2/l$.

In order to assess the proposals in a realistic circuit QED setup, we made numerical simulations for the Heisenberg dynamics between two qubits in the transmon regime coupled to a stripline waveguide resonator. 
We estimate the influence on the proposal of a state-of-the-art XY dynamics, given as an effective dispersive Hamiltonian, obtained at second order from the first order one,
\begin{equation}
H_{\textrm t}=\sum_{i=0}^2\sum_{j=1}^2\omega_i^j \ket{i,j}\bra{i,j}+\omega_ra^{\dagger}a+\sum_{i=0}^2\sum_{j=1}^2 g_{i,i+1}(\ket{i,j}\bra{i+1,j}+\textrm{H.c.})(a+a^{\dagger}).\label{HamcircuitQED}
\end{equation}

Here, $\omega_r$ is the resonance frequency of the resonator, and $\omega_i^j$ is the transition frequency of the $i$th level, with respect to the ground state, of the $j$th transmon qubit. We take into account the first three levels for each qubit, and an anharmonicity factor given by $\alpha_r=(\omega_2^j-2\omega_1^j)/\omega_1^j=-0.1$, standard for transmon qubits~\cite{Koch2007}. We consider equal transmons with frequencies $\omega_1^{1,2}\equiv \omega_1=2\pi\times5$~GHz. The frequency of the resonator is fixed to $\omega_r=2\pi\times7.5$~GHz. We take into account the coupling strength between the different levels of a single transmon qubit~\cite{Koch2007} $g_{i,i+1}=\sqrt{i+1}g_0$, being $g_0=2\beta eV_{\textrm{rms}}=2\pi\times200$~MHz. The experimental parameters we consider are standard for circuit QED platforms and they can be optimized for each specific platform. The transmon-resonator Hamiltonian, in interaction picture with respect to the free energy $\sum_{i,j}\omega_i^j \ket{i,j}\bra{i,j}+\omega_ra^{\dagger}a$, produces an effective interaction between the first two levels of the two qubits $H_{\textrm{eff}}=[g_{01}^2\omega_1/(\omega_1^2-\omega_r^2)](\sigma_1^x\sigma_2^x+\sigma_1^y\sigma_2^y)$, where we have neglected the cavity population $\langle a^{\dagger}a\rangle\approx0$ and renormalized the qubit energies to include Lamb shifts. Here, we have considered the set of Pauli matrices for the subspace spanned by the first two levels of each transmon qubit, e.g. $\sigma_{1(2)}^x\equiv\ket{0,1(2)}\bra{1,1(2)}+\textrm{H.c.}$

\begin{figure}[t]
\center
\includegraphics[scale=0.2]{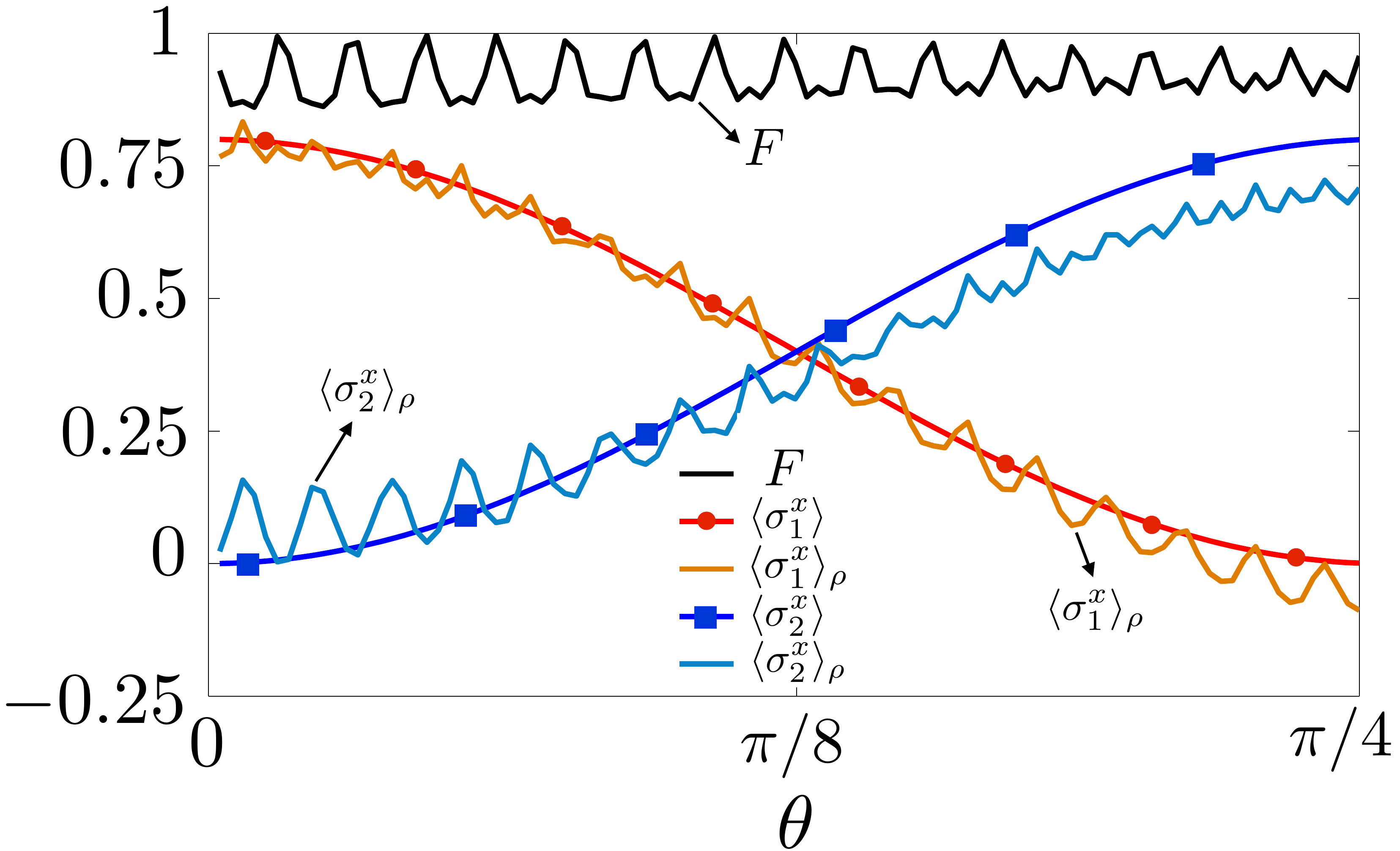}
\caption{Evolution of the emulated Heisenberg Hamiltonian for two superconducting transmon qubits, initialized in state $1/\sqrt{5}(\ket{\uparrow}+2\ket{\downarrow})\otimes\ket{\downarrow}$. The fidelity $F={\mathrm Tr}(\rho|\Psi_I\rangle\langle\Psi_I|)$ represents the performance of the protocol for the simulated phase $\theta$. The ideal spin evolution $\langle\sigma_i^x\rangle$ for both transmon qubits is depicted versus average values $\langle\sigma_{i}^x\rangle_\rho$ which are given through the qubit Hamiltonian $H_t$. Reprinted with permission from \cite{Urtzi14}, Copyright (2014) American Physical Society.\label{NumSim}} 
\end{figure}

In order to analyze the influence of decoherence in a state-of-the-art circuit QED setup, we compute the master equation evolution,
\begin{equation}
\dot{\rho}=-i[H_{\textrm{t}},\rho]+\kappa L(a)\rho+\sum_{i=1}^2\left(\Gamma_\phi L(\sigma^z_i)\rho+\Gamma_- L(\sigma_i^-)\rho\right),\label{master}
\end{equation}
where we define the Lindblad operators $L(\hat{A})\rho=(2\hat{A}\rho \hat{A}^{\dagger}-\hat{A}^{\dagger}\hat{A}\rho-\rho \hat{A}^{\dagger}\hat{A})/2$. We consider a damping rate for the cavity of $\kappa=2\pi\times10$~kHz, and a decoherence and decay rate for a single transmon qubit of $\Gamma_\phi=\Gamma_-=2\pi\times20$~kHz.  We compute a numerical simulation for the XYZ dynamics for two transmon qubits, following the scheme as in Fig.~\ref{Fig1Heis}a, using for the XY exchange gate steps the outcome of the evolution obtained from Eq.~(\ref{master}), and perfect single-qubit gates. We plot our result in Fig.~\ref{NumSim}. The dynamics for the density operator $\rho$, encoding the evolution of the two transmon qubits, is contrasted to the ideal quantum dynamics $|\Psi\rangle_I$, evolving with the Hamiltonian in Eq.~(\ref{XYZinhom}), where $J=g_{01}^2\omega_1/(\omega_1^2-\omega_r^2)\approx2\pi\times6$~MHz. It can be appreciated that the simulation fidelities $F={\mathrm Tr}(\rho|\Psi_I\rangle\langle\Psi_I|)$ obtained are good for nontrivial evolutions. We point out that the application of the XYZ Hamiltonian on an initial state, corresponding to an eigenstate of the ZZ operator, would be just equivalent to the one of the XY exchange dynamics. To show characteristic behavior of the XYZ interaction, we considered an initial state which does not have this feature. One can as well appreciate the standard short-time fidelity fluctuations  due to the spurious terms of the dispersive exchange Hamiltonian. Making use of a larger detuning of the qubits from the cavity, the contribution of the non-dispersive part of the interaction can be reduced, increasing the total protocol fidelity.  

Summarizing, we have proposed a digital quantum simulation of spin systems with circuit QED platforms. We have analyzed a prototypical model: the Heisenberg interaction. Moreover, we have studied the feasibility of the protocol with current technology of transmon qubits coupled to microwave cavities. These protocols may be generalized to many-body spin systems, paving the way towards universal digital quantum simulation of spin models with superconducting circuits. 

\subsection{Digital Quantum Simulations of Quantum Field Theories with Superconducting Circuits}
\label{subsec:dig2}

Our current knowledge of the most fundamental processes in the physical world is based on the framework of interacting quantum field theories~\cite{Peskin}. In this context, models involving the coupling of fermions and bosons play a prominent role. In these systems, it is possible to analyze fermion-fermion scattering mediated by bosons, fermionic self-interactions, and bosonic polarization. In this section, we will study~\cite{Laura14} a quantum field theory model with the following assumptions: (i) 1+1 dimensions, (ii) scalar fermions and bosons, and described by the Hamiltonian ($\hbar=c=1$)
\begin{eqnarray} \label{complex}
H = \int dp \ \omega_p (b^{\dag}_{p}b_{p} + d^{\dag}_{p}d_{p}) + \int dk \ \omega_k a^{\dag}_ka_k +  \int dx \ \psi^{\dag}(x)\psi(x)A(x).
\end{eqnarray}
Here, $A(x)=i\int dk \ \lambda_k \sqrt{\omega_{k}/4\pi}( a^{\dag}_k e^{-i k x} - a_k  e^{i k x} )$ is a bosonic operator, with coupling constants $\lambda_k$, and $\psi(x)$ is a fermionic field, $b^{\dagger}_p$($b_p$) and $d^{\dagger}_p$($d_p$) are its corresponding fermionic and antifermionic creation(annihilation) mode operators for frequency $\omega_p$, while $a^{\dagger}_k$($a_k$) is the creation(annihilation) bosonic mode operator associated with the frequency $\omega_k$. We propose a protocol for the scalable and efficient digital-analog quantum simulation of interacting fermions and bosons, based on Eq. (\ref{complex}), making use of the state-of-the-art circuit QED platforms. In this fast-evolving quantum technology, one has the possibility of a strong coupling of artificial atoms with a one-dimensional bosonic continuum.

In order to map the proposed model to the circuit QED setup, we consider a further assumption in Eq. (\ref{complex}): (iii) one fermionic and one antifermionic field modes~\cite{Casanova12} that interact via a bosonic continuum. Accordingly, the interaction Hamiltonian is given by
\begin{eqnarray}
H_{\rm int} &=& i \int dx dk\lambda_k\sqrt{\frac{\omega_{k}}{2}} \ \Big( |\Lambda_1(p_{f},x,t)|^{2} b^\dag_{\rm in}b_{\rm in}  + \Lambda_1^*(p_{f},x,t)\Lambda_2(p_{\bar{f}},x,t) b^\dag_{\rm in}d^\dag_{\rm in} \label{simple} \\ 
&& + \Lambda_2^*(p_{\bar{f}},x,t)\Lambda_1(p_{f},x,t) d_{\rm in}b_{\rm in}  + |\Lambda_2(p_{\bar{f}},x,t)|^{2} d_{\rm in}d^\dag_{\rm in}\Big) \left( a^{\dag}_k e^{-i k x} - a_k  e^{i k x} \right).
\nonumber
\end{eqnarray}
The fermionic and antifermionic creation and annihilation operators obey anticommutation relations $\{b_{\rm in},b^\dag_{\rm in}\}=\{d_{\rm in},d^\dag_{\rm in}\} = 1$, and the bosonic creation and annihilation operators satisfy commutation relations $[a_k,a^{\dag}_{k^{\prime}}]\!=\!\delta(k-k')$. 
Here, we have spanned the field $\psi(x)$ in terms of two comoving anticommuting modes as a first order approximation, neglecting the remaining anticommuting modes. These are given by the expressions,
\begin{eqnarray}
b^{\dag}_{\rm in} &=& \int dp \ \Omega_f(p_{f},p)b^{\dag}_pe^{-i\omega_p t} \label{comovingb} \\
d^{\dag}_{\rm in} &=& \int dp \ \Omega_{\bar{f}}(p_{\bar{f}},p)d^{\dag}_pe^{-i\omega_p t} \label{comovingd} ,
\end{eqnarray}
where $\Omega_{f,\bar{f}}(p_{f,\bar{f}},p)$ are the fermion and antifermion wavepacket envelopes with average momenta $p_f$ and $p_{\bar{f}}$, respectively. 

Thus, the fermionic field reads
\begin{equation}\label{coferm}
\psi(x) \simeq \Lambda_1(p_{f},x,t)b_{\rm in}+\Lambda_2(p_{\bar{f}},x,t)d^{\dag}_{\rm in} ,
\end{equation}
where the coefficients can be computed by considering the anticommutators $\{\psi(x),b^{\dag}_{\rm in}\}$ and $\{\psi(x),d_{\rm in}\}$ as follows
\begin{eqnarray}
\Lambda_1(p_{f},x,t) &=& \{\psi(x),b^{\dag}_{\rm in}\} = \frac{1}{\sqrt{2\pi}} \int \frac{dp}{\sqrt{2\omega_{p}}} \Omega(p_{f},p)e^{i(px-\omega_p t)}\label{cb} , \\ 
\Lambda_2(p_{\bar{f}},x,t) &=& \{\psi(x),d_{\rm in}\} = \frac{1}{\sqrt{2\pi}} \int \frac{dp}{\sqrt{2\omega_{p}}} \Omega(p_{\bar{f}},p)e^{-i(px-\omega_p t)}\label{cd} ,
\end{eqnarray}
where we have considered $\psi(x)$ in the Schr\"{o}dinger picture. 

With this proposal, we think that emulating the physics of a discrete number of fermionic field modes coupled to a continuum of bosonic field modes will significantly enhance the quantum simulations of full-fledged quantum field theories.

\begin{figure*}[t]
\includegraphics[width=\textwidth]{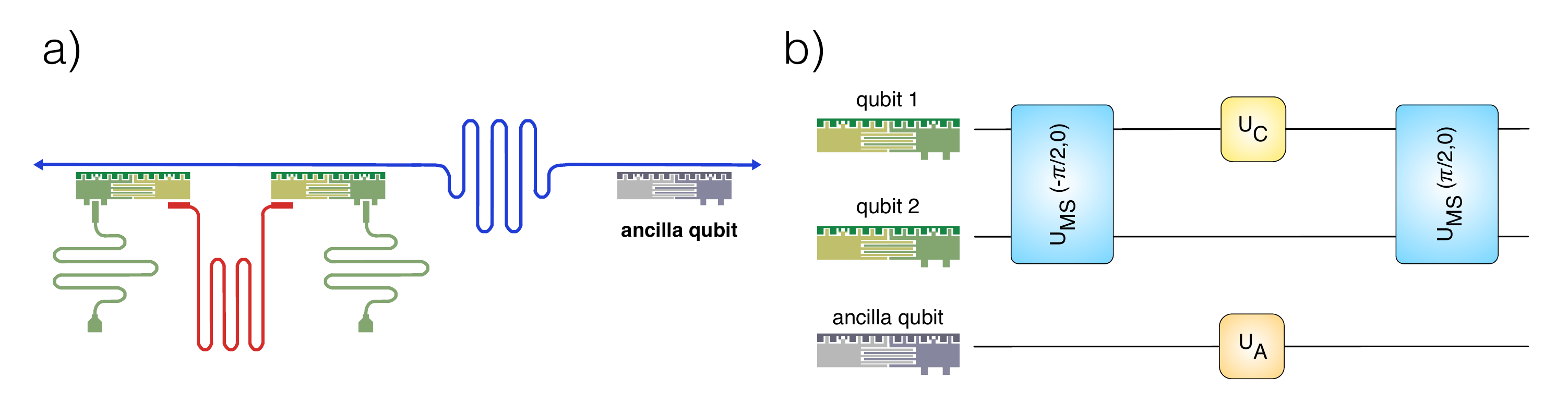}
\caption{\label{Fig1QFT}  a) Scheme of our protocol for emulating fermion-fermion scattering in QFTs. An open line supporting a bosonic continuum is coupled to three superconducting transmon qubits. A second, one-dimensional stripline waveguide, forming a cavity, contains a single bosonic mode of the microwave field and couples with two superconducting transmon qubits. Each of the qubits can be locally addressed through external flux drivings generating fluxes $\Phi^j_{\rm ext}$ and $\bar{\Phi}^j_{\rm ext}$ to adjust the coupling and its associated frequencies. b) Sequential protocol of multiple and single qubit gates, in a single digital step, acting on transmon qubits to produce two-qubit gates interacting with the continuum. Reprinted with permission from \cite{Laura14}, Copyright (2015) American Physical Society.}
\end{figure*}

We now use the Jordan-Wigner transformation~\cite{JordanWigner,AltlandSimons} that maps fermionic mode operators onto tensor products of spin operators: $b^\dag_l=\prod_{r=1}^{l-1}\sigma_l^-\sigma_{r}^z$, and $d^\dag_m=\prod_{r=1}^{m-1}\sigma_m^-\sigma_{r}^z$, where $l=1,2,...,N/2$,  $m=N/2+1,...,N$, with $N$  the total number of fermionic and antifermionic modes. Thus, Hamiltonian~(\ref{simple}) is reduced to just three different kinds of couplings: single and two-qubit gates interacting with the bosonic continuum $H_1= i\sigma_{j}\int dx dk~g_k(a_k^{\dag} e^{-ikx}-a_k e^{ikx})$, $H_2= i(\sigma_j\otimes\sigma_{\ell})\int dx dk~g_k(a_k^{\dag} e^{-ikx}-a_k e^{ikx})$, with $\sigma_{q} = \{\sigma_x, \sigma_y, \sigma_z\}$ for $q=1, 2, 3$, and couplings that involve only bosonic field modes, $H_3= i\int dx dk~g_k(a_k^{\dag} e^{-ikx}-a_k e^{ikx})$. Therefore, the quantum simulator should produce a way of generating multiqubit entangling gates and coupling qubit operators to a bosonic continuum through a digital-analog method~\cite{Casanova12}.

Circuit QED platforms consisting of the coupling between coplanar waveguides (CPW) and transmon qubits~\cite{Gambetta2011,Houck2011,Steffen2013} are an appropriate setup to implement our digital-analog simulator model. We depict in Fig.~\ref{Fig1QFT}a a scheme of our setup, which is based on a microwave transmission line supporting a continuum of electromagnetic field modes (open line) that interacts with three qubits in the transmon regime. Moreover, we consider a microwave stripline resonator with a single bosonic mode coupled only with two of the qubits. We point out that two superconducting transmon qubits may interact at the same time with both CPWs, while the ancilla transmon qubit will interact only with the open transmission line. 

\begin{figure*}[t]
\includegraphics[width=\textwidth]{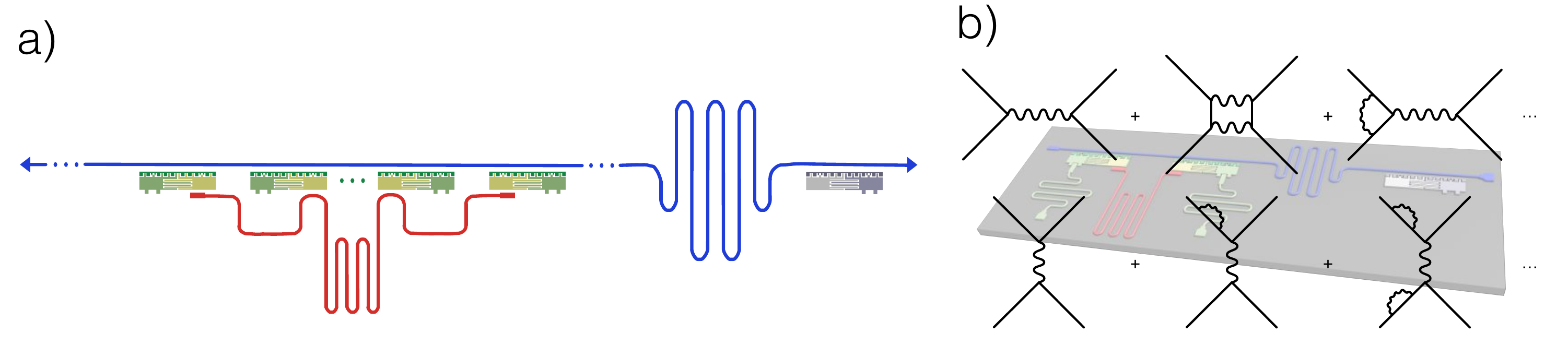}
\caption{\label{Fig2QFT} a)  Schematic representation for the realization of $N$ fermionic field modes interacting with a bosonic continuum. Each fermionic field mode is mapped onto a nonlocal spin operator implemented among $N$ superconducting transmon qubits. b) Feynman diagrams related to the quantum simulation of two fermionic field modes interacting with a bosonic continuum in a circuit QED setup, as described in the text. Reprinted with permission from \cite{Laura14}, Copyright (2015) American Physical Society.}
\end{figure*}

In our proposal~\cite{Laura14}, we take into account tunable couplings among each transmon qubit and the CPWs, as well as tunable transmon qubit energies via applied magnetic fluxes. More specifically, our method for emulating fermion-fermion scattering will be based on the capacity to turn on/off each CPW-qubit coupling with tunable parameters. The latter may be performed by using controlable coupling superconducting qubits,~\cite{Gambetta2011,Houck2011} and typical techniques of band-pass filter~\cite{Pozar} to apply in the open line, in the sense that just a finite bandwidth of bosonic field modes plays a role in the evolution. In this respect, to decouple a superconducting qubit from the open transmission line may be achieved by shifting the qubit frequency outside of the permitted bandwidth. Moreover, our proposal may be extrapolated to many fermionic field modes by considering more superconducting transmon qubits, as shown in Fig.~\ref{Fig2QFT}.

In our proposal, the transmon qubit-continuum and the transmon qubit-resonator couplings are expressed through the interaction Hamiltonian
\begin{eqnarray}
\label{hsetup}
H_{\rm int} = && \,\, i \sum^3_{j=1}\sigma^{y}_j\int dk~ \beta(\Phi^j_{\rm ext},\bar{\Phi}^j_{\rm ext}) g_k(a^{\dag}_ke^{-ikx_j} - a_ke^{ikx_j})\nonumber \\
&& + \, i \sum^2_{j=1}\alpha(\Phi^j_{\rm ext},\bar{\Phi}^j_{\rm ext})g_j\sigma^{y}_j(b^{\dag}-b),
\end{eqnarray}
where $a^{\dagger}_k$($a_k$)  and $\omega_k$ are the creation(annihilation) operator and the free energy associated with the $k$th continuum field mode, respectively. In addition, $b^{\dagger}$($b$) denotes the creation(annihilation) bosonic operator in the microwave cavity, and $\sigma^y$ is the corresponding Pauli operator. The couplings $g_k$ and $g_j$ are a function of specific properties of the CPW as for example the photon frequencies and its impedance. Moreover, $x_j$ denotes the $j$th transmon qubit position, and the function $\beta(\alpha)$ can be changed over the interval $[0,\beta_{\rm{max}}]([0,\alpha_{\rm{max}}])$ via applied magnetic fluxes $\Phi^j_{\rm ext}$ and $\bar{\Phi}^j_{\rm ext}$, which are externally driven on the $j$th superconducting qubit. We point out that these external magnetic fluxes allow as well to modify the qubit frequency.  

We show now how the interaction Hamiltonian (\ref{hsetup}) can emulate the evolution associated with Hamiltonian~(\ref{simple}). We plot in Fig.~\ref{Fig1QFT}b the quantum gates needed for emulating two-qubit operations interacting with the continuum in a single digital step~\cite{Lloyd96,Casanova12} to be performed by  our digital-analog emulator. In this proposal, making use of a  superconducting circuit framework, each unitary operator will be associated with the dynamics under the Hamiltonian~(\ref{hsetup}) for corresponding external fluxes $\Phi^j_{\rm ext}$ and $\bar{\Phi}^j_{\rm ext}$.  More concretely, the operators acting on the first two superconducting transmon qubits are, sequentially applied,  a M\o lmer-S\o rensen~\cite{Molmer1998} operator $U_{\rm MS}(\pi/2, 0)$ performed through the cavity~\cite{Mezzacapo2014}, a local rotation $U_C = \exp[-\phi\sigma_1^y \int dk~g_k(a_k^{\dag} e^{-ikx}-a_k e^{ikx})]$ that produces an interaction between the spin matrices and the bosonic field continuum, and the inverse  M\o lmer-S\o rensen operator $U_{\rm MS}(-\pi/2, 0)$. The combination of the three gates  will produce  the corresponding two-qubit gate coupled with the continuum of bosonic modes, $H_2= i(\sigma_j\otimes\sigma_{\ell} )\int dk~g_k(a_k^{\dag} e^{-ikx}-a_k e^{ikx})$. 

The $U_{c}$ operator will be employed independently on each superconducting transmon qubit to produce the corresponding single-qubit gates interacting with the bosonic continuum. Moreover, the auxiliary transmon qubit permits to produce the operators involving only the bosonic field modes by using a gate,
\begin{equation}
U_A = \exp[-\phi\sigma_A^z \int dk g_k (a_k^{\dag} e^{-ikx}-a_k e^{ikx})], 
\end{equation}
where $\sigma_A^z$ is the corresponding Pauli operator. The necessary operator is achieved by initializing the auxiliary qubit in an eigenstate of $\sigma^z_A$. An equivalent sequence of operators can be applied on more superconducting qubits for scaling the model in order to emulate couplings involving many fermionic field modes. 

The path for scaling this proposal to many fermionic field modes is to take into account more superconducting transmon qubits interacting both with the resonator and with the open line, as we plot in Fig. \ref{Fig2QFT}. When one considers $N$ superconducting qubits, $N$ fermionic field modes can be emulated. Therefore, our protocol can realize a large number of fermionic field modes coupled to the bosonic field continuum. This proposal will represent a significant step forward towards an advanced quantum simulation of full-fledged quantum field theories in controllable superconducting qubit setups.

Making use of the proposed method, one can extract information of relevant features of quantum field theories, as for example pair creation and annihilation of fermions as well as self-interaction, mediated via a bosonic field continuum. This quantum simulation is based on unitary gates related to Hamiltonian~(\ref{simple}). In this respect, as opposed to standard perturbative techniques in quantum field theories, the realization of our proposal will be associated with an infinite number of Feynman diagrams and a finite number of fermionic field modes. Accordingly, this path towards full-fledged QFTs is at variance from standard techniques, because it needs the addition of more fermionic field modes instead of more perturbative Feynman diagrams. On the other hand, the fact that we consider a continuum of bosonic field modes in circuit QED makes our protocol nearer to the targeted theory.

To conclude, we have introduced a method for a digital-analog quantum emulation of fermion-fermion scattering and quantum field theories with circuit QED. This quantum platform benefits from strong coupling between superconducting transmon qubits with a microwave cavity and a continuum of bosonic field modes. Our method is a significant step forward towards efficient quantum simulations of quantum field theories in perturbative and nonperturbative scenarios.

\section{Conclusion}
\label{sec:Conc}

We have presented the topic of analog and digital quantum simulations in the light of circuit QED technologies. In particular, we have discussed the basic concepts of circuit network theory and their applications to electric circuits operating at the quantum degeneracy regime imposed by the superconducting state.  

We have shown how circuit QED with a transmon or a flux qubit represents a building block for circuit QED lattices aiming at simulating Hamiltonians of condensed matter physics such us the Bose-Hubbard model, the Jaynes-Cummings-Hubbard model, models with nearest-neighbor Kerr nonlinearities that exhibit Bose-Hubbard features, the bosonic Kagome lattice, and the Rabi-Hubbard model. The latter has interesting predictions provided by the counter-rotating terms that appear in the system Hamiltonian.

Regarding the digital approach of quantum simulations, we have presented two recent developments, i.e., the simulation of spin systems, and the digital/analog simulation of quantum field theories exemplified by the fermion-fermion scattering mediated by a continuum of bosonic modes. These theoretical efforts are based on the state-of-the-art in circuit QED with transmon qubits, and may pave the way for experimental developments in the near future.

\section*{Acknowledgments}
\label{sec:Ack}

We acknowledge funding from the Basque Government IT472-10; Ram\'on y Cajal Grant RYC-2012-11391; Spanish MINECO FIS2012-36673-C03-02 and FIS2015-69983-P; UPV/EHU UFI 11/55; UPV/EHU Project EHUA14/04; the Fondo Nacional de Desarrollo Cient\'ifico y Tecnol\'ogico (FONDECYT, Chile) under grant No. 1150653; PROMISCE and SCALEQIT EU projects.

\end{document}